\documentclass[]{emulateapj}
\usepackage{amssymb,amsmath}

\shorttitle{Fluorescent Ly$\alpha$ emission}
\shortauthors{Cantalupo et al.}

\begin{document}
\def\simlt{\mathrel{\rlap{\lower 3pt\hbox{$\sim$}} \raise
        2.0pt\hbox{$<$}}} \def\simgt{\mathrel{\rlap{\lower
        3pt\hbox{$\sim$}} \raise 2.0pt\hbox{$>$}}}
\def\simgt{\mathrel{\rlap{\lower 3pt\hbox{$\sim$}} \raise
        2.0pt\hbox{$>$}}} \def\simgt{\mathrel{\rlap{\lower
        3pt\hbox{$\sim$}} \raise 2.0pt\hbox{$>$}}}

\title
{Fluorescent Ly$\alpha$ emission from the high-redshift 
intergalactic medium}

\author{Sebastiano Cantalupo, Cristiano Porciani, Simon J. Lilly and Francesco Miniati}
\affil{Institute for Astronomy, ETH Z\"urich, CH-8093 Z\"urich, Switzerland
\\
cantalupo@phys.ethz.ch, porciani@phys.ethz.ch, lilly@phys.ethz.ch, fm@phys.ethz.ch}


\begin{abstract}
We combine a high-resolution hydro-simulation of the $\Lambda$CDM 
cosmology with two radiative
transfer schemes (for continuum and line radiation) 
to predict the properties, spectra and spatial distribution 
of fluorescent Ly$\alpha$ emission at $z\sim3$.
We focus on line radiation produced by recombinations in the dense
intergalactic medium ionized by UV photons.
In particular,
we consider both a uniform background and the case where gas clouds
are illuminated by a nearby quasar.
We find that
the emission from optically thick regions is substantially
less than predicted from the widely used static, plane-parallel model.
The effects induced by a realistic velocity field 
and by the complex geometric structure of the emitting regions
are discussed in detail.
We make predictions for the expected brightness and size distributions 
of the fluorescent sources.
Our results account for recent null detections 
and can be used to 
plan new observational campaigns both in the field 
(to measure the intensity of the diffuse UV background)
and in the proximity of bright quasars 
(to understand the origin of high colum-density absorbers).

\end{abstract}


\section{Introduction}
Hydrogen absorption-line systems observed shortward of Ly$\alpha$ 
emission in quasar 
spectra constitute an important probe of the physical state of the 
intergalactic medium at high-redshift.
These spectral features are shaped by
the combined action of gravity, hydrodynamics and photoionization processes
which determine the local density and the velocity field 
of neutral hydrogen within the absorbers.
Numerical simulations suggest that the so called Lyman-$\alpha$ forest
is generated by diffuse, sheetlike and filamentary structures with a mean
density which is between 1 and 10 times higher than the cosmic average
(Cen et al. 1994; Zhang, Anninos \& Norman 1995; Hernquist et al. 1996;
Miralda-Escud\'e et al. 1996).
These low-column-density systems
are highly ionized by the extragalactic background of Lyman continuum photons
generated by young stellar populations and quasars.
At the opposite extreme,
Lyman-limit (LLS, $N_{\rm HI}> 10^{17.2} \ {\rm cm}^{-2}$)
and damped Lyman-$\alpha$ (DLA, $N_{\rm HI}> 10^{20.3} \ {\rm cm}^{-2}$) 
systems  
correspond to concentrations of atomic hydrogen which are optically
thick to the cosmic ionizing background.
Numerical simulations suggest that they arise in dense gas clouds with 
a meatball topology. On cosmological scales, they appear to form  
a collection of isolated clouds which trace the cosmic web.

Optically thick clouds are expected to emit
fluorescent Ly$\alpha$ photons produced in hydrogen recombinations
(Hogan \& Weymann 1987; Gould \& Weinberg 1996).
This emission is concentrated in the outer parts of the clouds where
hydrogen is significantly ionized by the external UV background
($\tau_{\rm LL}\sim 1$).
However,
Ly$\alpha$ photons cannot directly escape the clouds because of the large 
optical depth in the center of the line 
($\tau_{{\rm Ly}\alpha} \simeq 10^{4} \tau_{\rm LL}$).
Each photon thus suffers a large number of resonant scatterings
(more precisely: absorptions and re-emissions) 
by neutral hydrogen atoms in the ground state.
Each scattering adds a small Doppler shift to the frequencies of the photons 
due to the thermal (and turbulent) motions of the atoms.
Therefore, photons execute a random walk both in frequency and in physical 
space until their frequencies are shifted sufficiently away from the line 
center and they are able to escape the medium in a single flight 
(Zanstra 1949). 

Monte Carlo simulation (e.g. Ahn, Lee \& Lee 2001; Zheng \& Miralda-Escud\'e 2002b 
and references therein) is the most popular method for addressing the
radiative transfer problem. 
Analytical solutions only exist for highly symmetric systems.
For instance, the emerging spectrum from a plane-parallel and static
homogeneous slab
is characterized by two sharp peaks in the Doppler wings of the line   
(Neufeld 1990 and references therein).
The plane-parallel solution
approximately holds also for self-shielded systems where 
the ionized layer which surrounds the neutral region is
thin with respect to the characteristic radius of the cloud.
In this ideal case, optically thick systems act as efficient mirrors
which convert nearly $60\%$ of the impinging ionizing flux into Ly$\alpha$ 
photons (Gould \& Weinberg 1996). 

Direct imaging of fluorescent sources
would lead to a major advance in
our understanding of galaxy formation.  
Determining the size distribution of LLS at $z\simgt 3$ 
would be crucial to distinguish
whether they arise from photoionized clouds in galactic halos 
(Steidel et al. 1995; Mo \& Miralda-Escud\'e 1996) or in minihaloes formed prior
to reionization (Abel \& Mo 1998).
At the same time, 
the intensity of the cosmic UV background could be inferred
from the observed brightness of the fluorescent emission.

With present-day technology, the detection of fluorescent  
emission from high-redshift gas condensations is challenging, but not 
impossible.
At $z\sim 3$, the intensity of the diffuse ionizing background 
(e.g. Haardt \& Madau 1996) corresponds to a Ly$\alpha$ 
surface brightness of the order of
$10^{-20}
\mathrm{erg}\,\,\mathrm{cm}^{-2}\,\mathrm{s}^{-1}\,\mathrm{arcsec}^{-2}$.
It is then not surprising that blind searches have only produced a number
of null results 
(Lowenthal et al. 1990; Mart\'inez-Gonzalez et al. 1995;
Bunker, Marleau \& Graham 1998).
Positive fluctuations in the ionizing background can be used
to increase the signal. For instance, clouds lying close to a bright
quasar are exposed to a stronger UV flux (with respect to an ``average'' cloud)
and are then expected to be brighter in fluorescent Ly$\alpha$.
Very recently, 
Francis \& Bland-Hawthorn (2004) presented a deep narrow-band search
for Ly$\alpha$ emission in a field which lies next to the quasar PKS 0424-131.
Based on quasar-absorption-line statistics and on simple models for 
fluorescent emission (Gould \& Weinberg 1996), they expected to detect
more than 6 clouds
but none were seen. These null results highlight the need for a more
sophisticated analysis of fluorescent Ly$\alpha$ emission in realistic
environments.

In this paper, we present accurate models of the fluorescent Ly$\alpha$ 
emission from LLSs at redshift $z\sim 3$. Our study proceeds
in three steps. 
First, we perform a hydrodynamical simulation of structure formation to
compute the cosmological distribution of the baryons at $z=3$. 
A simple radiative transfer scheme is then used to propagate the ionizing
radiation through the computational box and to compute the distribution of 
neutral hydrogen and of recombinations.
Finally, a three-dimensional Monte Carlo code is used 
to follow the transfer of Ly$\alpha$ photons.
As ionizing radiation,
we first consider the diffuse background generated by the UV emission 
of galaxies and quasars (Haardt \& Madau 1996).
We then discuss an inhomogeneous case where
the ionizing flux from a quasar 
(which lies in the foreground of the gas clouds)
is superimposed to the uniform background.
Our detailed numerical analysis
shows that simplified models (e.g. Gould \& Weinberg 1996)
tend to overpredict the Ly$\alpha$ flux emitted from optically thick regions. 

The structure of the paper is as follows.
We describe our numerical techniques in \S 2 and present our results
in \S 3 where we also discuss the implications of our analysis for
present and future observations. 
Finally, we discuss the limitations of our approach in \S 4 
and we conclude in \S 5.

\section{Method}
\subsection{Cosmological simulation}\label{CS}

The formation and evolution of the large-scale structure in a ``concordance''
$\Lambda$CDM cosmological model is followed
by means of an Eulerian, grid based Total-Variation-Diminishing
hydro+N-body code (Ryu et al. 1993).
We assume that 
the mass density parameter $\Omega_0=0.3$
(with a baryonic contribution  $\Omega_{\rm b}=0.04$),
the vacuum-energy density parameter $\Omega_\Lambda= 1- \Omega_m= 0.7$
and the present-day value of the Hubble constant
constant $H_0=100\,h $ km s$^{-1}$ Mpc$^{-1}$
with $h=0.67$.
The simulation is started at redshift $z=60$
and follows the evolution of Gaussian density fluctuations 
characterized by a primordial 
spectral index $n=1$ and ``cluster-normalization'' $\sigma_8=0.9$
(with $\sigma_8$ the rms linear density fluctuation within a sphere
with a comoving radius of $8\, h^{-1}$ Mpc).
This is consistent with the most recent joint analyses of temperature
anisotropies in the cosmic microwave background and galaxy clustering
(e.g. Tegmark et al. 2004 and references therein).
We use a comoving computational box size of $10\,h^{-1}\,$Mpc where
the dark matter distribution is traced by 256$^3$ particles and
the gas component is evolved on a comoving grid with 512$^3$ zones. 
The nominal spatial resolution for the gas (the mesh size) is  
$\sim20\,h^{-1}\,$kpc (comoving) with the mean baryonic mass
in a cell being $\sim 10^5\,h^{-1}\,M_\odot$.
On the other hand,
each dark matter particle has a mass of $5\times 10^6\,h^{-1}\,M_\odot$.
All the results presented in this work are derived from the
$z=3$ output of a simulation which does not include 
radiative cooling of the gas.
The limitations of this assumption are briefly discussed in \S \ref{discus}.
We defer a detailed analysis of the radiative case to future work.


\subsection{Radiative transfer of UV radiation}\label{RTran}

In order to compute the distribution of neutral hydrogen within a
snapshot of the computational box,
we need to simultaneously solve   
the radiative transfer problem for UV radiation and
the rate equations describing the balance between the ionization and 
recombination rates.

For simplicity, we assume that hydrogen is in ionization equilibrium and use 
the ``on the spot" approximation (Baker 1962): 
\begin{equation}\label{csi_ion}
(1-x)\,n_{\rm H}\,\int_{\nu_0}^{\nu_{\rm up}}\!\!\!\!\!{\mathrm d}\nu 
\frac{ \sigma_{\nu}}{h_{\rm P}\nu} \int_{4 \pi} \!\!\!\!{\mathrm d}\Omega\,J_{\nu} (\Omega)\,
= x\, n_{\rm H}\,n_{\rm e}\,\alpha_{\rm B}(T)
\end{equation}
where $h_{\rm P}$, $n_{\rm e}$,
$x$, $n_{\rm H}$, $\sigma_{\nu}$, $T$ and $\alpha_{\rm B}$ respectively denote the Planck constant,
the electron number density and
the hydrogen ionized fraction, volume number density, ionization cross section,  temperature and case B recombination coefficient (for which we use the fit by Hui \& Gnedin 1997). The intensity of ionizing radiation per unit frequency and solid angle is given by
$J_{\nu}$  (in erg $\mathrm{cm^{-2}}$ $\mathrm{s^{-1}}$ $\mathrm{sr^{-1}}\mathrm{Hz^{-1}}$).
The frequency integral in equation (\ref{csi_ion}) extends from the hydrogen ionization threshold,
$h_{\rm P}\nu_0$=13.6 eV, to a maximum frequency  $\nu_{\rm up}$ (which is, formally,
infinite).
A good approximation for our purposes is to assume 
$\nu_{\rm up}=4\,\nu_0$,
(i.e. set the intensity of radiation to zero at frequencies above the 
ionization threshold for HeII).
The motivation is twofold.
First, nearly all the photons with $\nu>4\,\nu_0$ (which anyway contribute only a few per cent of the
energy available for H ionization in the UV background) 
will be absorbed by He atoms (Haardt \& Madau 1996).
Second, HeII recombines faster than HI and the 
intensity of radiation at the HeII Lyman limit is typically lower than at $\nu_0$. Therefore, HeII is 
more easily shielded from the ionizing background with respect to HI (Miralda-Escud\'e \& Ostriker 1990).
This implies that HeII-ionizing photons are absorbed
in the outer regions of the gas concentrations where H is nearly fully ionized. 
In order to describe the hydrogen shielding layers we thus neglect HeIII 
and assume that the neutral fraction of 
He coincides with $1-x$ (Zheng \& Miralda-Escud\'e 2002a).
For a helium abundance of $Y=0.24$,
this corresponds to assuming $n_{\rm e}=\beta\,x\,n_{\rm H}$ with $\beta\simeq 1.08$
(see also \S \ref{fcf}).
Other than this, the presence of He atoms is neglected in equation (\ref{csi_ion}).
Given that HeII recombinations produce HI-ionizing photons and the relatively small number density of 
helium atoms
and ions, this approximation should be reasonably accurate.
Note that also recombination radiation from HeIII can ionize HI.
However, considered the different spatial distribution of HeIII and HI discussed above and 
the characteristic HeIII-recombination time scales, 
we neglect the small local corrections to the HI-ionizing background deriving from this effect.

In each cell of the simulation, the diffuse ionizing background is approximately described
by following the radiative transfer along 6 ``light-rays" which propagate parallel (and antiparallel)
 to the main axes of the computational box. 
With this numerical trick  we can treat anisotropic backgrounds (created, for instance, by
shadowing effects) with a minimal request of CPU time (see Appendix 
\ref{6dtest} for a test of this approximation).
Let us denote by $\tau_i(\nu)$ the optical depth of a given cell along the $i$-th ray.
This quantity is computed by integrating the product $(1-x)\,n_{\rm H}\,\sigma_\nu$ 
from a given starting location (a light source) in the box (see below) up to the first point
of the cell crossed by the ray. 
The closest face of the cell is then exposed to a radiation field with intensity 
$J_\nu^{\rm in}\,e^{-\tau_i(\nu)}$, where $J_{\nu,i}^{\rm in}$ 
denotes the input ionizing radiation before it is filtered by the gas distribution in the box.
Let us also indicate with $\Delta \tau(\nu)=(1-x)\,n_{\rm H}\,\sigma_\nu\,L$
(with $L$ the cell size in physical units)
the optical-depth variation within the cell measured along one of its principal axes.
In order to implement a photon conserving scheme, we replace the left-hand side in 
equation (\ref{csi_ion}) with the quantity
\begin{equation}
\label{RT}
\frac{4\pi}{6}\sum_{i=1}^{6} \int_{\nu_0}^{\nu_{\rm up}} 
\frac{J_{\nu,i}^{\rm in}}{h_{\rm P}\nu}\,e^{-\tau_i(\nu)}\,\frac{1-e^{-\Delta \tau(\nu)}}{L}
\end{equation}
where the sum is taken over the six rays (labeled by the index $i$).
This corresponds to the number of ionizing photons (per unit volume and time)
which are deposited in a given cell by the six rays. 
To describe the diffuse UV background, we assume that $J_{\nu,i}^{\rm in}=J_\nu^{\rm HM}$ 
with $J_\nu^{\rm HM}$ the intensity of radiation derived at $z=3$ by   
Haardt \& Madau (in preparation,
hereafter HM) considering the emission from observed quasars and galaxies 
after it is filtered through the Ly$\alpha$ forest
\footnote{This is obtained using the most recent
results regarding the quasar luminosity function and cosmic evolution within
a concordance cosmological model. It assumes that the galaxy escape fraction of ionizing
raduation is $f_{\rm esc}=0.1$ and that
the energy spectral index for quasar radiation is $\alpha=1.8$. 
The resulting
hydrogen ionization rate is a factor 1.16 smaller than in the models by 
Haardt \& Madau (1996) used by Gould \& Weinberg (1996). The spectrum is
available at {http://pitto.mib.infn.it/$\sim$haardt/refmodel.html}.}. 
We assume that
underdense cells are exposed to the full, isotropic  background.
On the other hand, overdense cells see an anisotropic
radiation field which is computed by using equation (\ref{RT}) to
propagate the input background starting from the surface where $\rho=\bar{\rho}$.
The intensity of radiation (and thus $x$) in each overdense cell depends
on the ionized fraction of the surrounding region.
To solve the non-local equations,
we start our calculations by assuming that the whole simulation box is optically thin
(i.e. it is exposed to the input radiation field)
and we iterate the radiative transfer and ionization-equilibrium calculations 
until convergence (within 1\%) is reached in each overdense cell.   

We use a similar approach to discuss the anisotropic radiation field generated 
by a quasar lying in the foreground of the simulation box along the 
observer's line of sight.
For simplicity, we assume that the quasar lies distant enough from the 
simulated region that its emission
can be modeled as a train of plane waves impinging onto a face of the 
simulation box.
We also assume that the quasar input spectrum is identical 
to that of the cosmic background. 
Given that $J_\nu^{\rm HM}$ is well described by a power-law of index -1.25 between
$\nu_0$ and $3\nu_0$, this is a sufficiently good approximation for our purposes
(see also the extensive discussion in \S \ \ref{uvb}).
We then write the quasar ionizing flux  
(erg $\mathrm{cm^{-2}}$ $\mathrm{s^{-1}}$ $\mathrm{Hz^{-1}}$) as 
$F_\nu=\pi\, b\, J_\nu^{\rm HM}\,\delta_{1i}$
with $\delta_{ij}$ the Kronecker symbol and $b$ a dimensionless constant.
This is equivalent to using 
$J_{\nu,i}^{\rm in}=1.5\,b\,J_\nu^{\rm HM}\,\delta_{1i}$ 
in equation (\ref{RT}).
In this case, we compute
the optical depth starting from the face of the simulation box 
which is first reached by quasar light (i.e. along the direction $i=1$).

A self-consistent calculation of the gas temperature requires a joint treatment of radiative transfer and hydrodynamics which is still beyond present-day computing capabilities.
Assuming that the photoionized gas is in thermal equilibrium, we find  that
$T\simeq1-3\times 10^4$ K
for the typical densities in the shielding layers 
($100\simlt \rho/\bar{\rho} \simlt300$).
However, shock heating can easily drive the gas temperature to $10^{5-7}$ K.
This is particularly important for the low-density regions 
($\rho\simlt 100 \,\bar{\rho}$)
where cooling processes are inefficient and the shocked material 
remains hot (Theuns et al. 1998).
In our analysis, we assume that $T=2\times 10^4$ K everywhere.
This is an excellent approximation for highly overdense regions ($\rho\simgt 100 \,\bar{\rho}$) 
where the cooling time is shorter than the Hubble time and the gas temperature rapidly approaches
the equilibrium solution (Theuns et al. 1998).
Anyway, since the recombination coefficient $\alpha_{\rm B}$ has only a weak dependence on $T$, fixing the temperature to $2\times 10^4$ K in the whole simulation box does not seriously affect our results.

Note that,
at $T=2\times 10^4\,K$, the 
hydrogen recombination timescale is 
$t_{\rm rec}=2.26\, (\bar{\rho}/\rho) \times 10^{10}$ yr. 
Ionization equilibrium will approximately hold only 
where $t_{\rm rec}$ is shorter
than the characteristic quasar lifetime 
($\sim 10^8$ yr, Porciani,
Magliocchetti \& Norberg 2004), i.e. for $\rho\simgt 200\, \bar{\rho}$.
At lower densities, our assumption of ionization equilibrium will
then overestimate the hydrogen ionized fraction.
This is not a problem for our study since, in the vicinity of a quasar,
the ionizing flux is strong enough to nearly completely ionize
the low-density intergalactic medium.
It is worth noticing, however, that
regions with $\rho< 200\, \bar{\rho}$ will emit
their recombination radiation after the quasar has switched off and will 
not be detectable in a survey centred onto a bright quasar.

\subsection{The clumping factor}\label{fcf}

Hydro-simulations have a finite spatial resolution and cannot describe the gas 
distribution on arbitrarily small scales.  
In other words, they provide a coarse grained
representation of the density field.
However, the hydrogen recombination rate scales proportionally to the square 
of the local (i.e. fine grained) number density and is sensitive to
small-scale inhomogeneities (clumpiness) within a simulation cell. 
In order to keep track of this discrepancy, 
we re-write the mean recombination rate within a cell as
\begin{equation}
\label{newrec}
x^2\,\beta\,\mathcal{C}\,\langle n_{\rm H} \rangle^2\,\alpha_{\rm B}(T)
\end{equation}
where the average is taken over a simulation cell and
\begin{equation}
\mathcal{C}=\frac{\langle n_{\rm H}^2 \rangle}{\langle n_{\rm H} \rangle^2}
\end{equation}
denotes the clumping factor of the gas (we assume that different atomic species and ions
have the same spatial distribution).
In principal, the latter quantity
can be estimated by comparing simulations with different resolutions
and consistent initial conditions.
We assume that $\mathcal{C}$ is constant everywhere and 
we fix its value by imposing that the number density (per unit redshift) of LLSs in our 
simulation matches the observational data (P\'eroux et al. 2003).
 \footnote{
Note that the spectral resolution of the observational data roughly corresponds to our box size. 
Therefore we can safely compute the hydrogen column density by integrating $n_{\rm HI}$ along 
the entire box.}
This normalization procedure, which requires $\beta\,\mathcal{C}\simeq6$, 
partially overcomes the limitations of our simulation 
(limited resolution and any missing physics).
Note that the observational data constrain the product $\beta\,\mathcal{C}$ 
so that there is no need to specify a priori the He ionization state
as discussed in \S \ref{RTran}.

\subsection{Ly$\alpha$ emission}

Using equation (\ref{newrec}),
we compute the hydrogen recombination rate in each cell of the simulation.
In order to convert this quantity into an emission rate for  Ly$\alpha$ 
photons,
we need to evaluate how many recombinations ultimately lead to a $^2P\to{^1S}$ 
transition.
For $T=2\times 10^4$ K, 
nearly 44\% of the atoms directly recombine to the ground level while 35\% 
of the remaining cases ultimately produce excited atoms in
the $^2S$ state which decays to $^1S$ via two-photon emission
(both fractions are weakly dependent on the gas temperature,
see e.g. Osterbrock 1989).  
Therefore, if the gas is optically thin to UV photons, 
only a fraction 
$\epsilon_{\rm thin}=\alpha^{\rm eff}_{^2P}/\alpha_{\rm A}\sim 0.36$ 
(where $\alpha^{\rm eff}_{^2P}$ and $\alpha_{\rm A}$ denote
the effective recombination coefficient to the $2P$ level and
the case A total recombination coefficient, respectively)
of the recombinations 
yield a Ly$\alpha$ photon. However, in the optically thick case, 
continuum photons 
produced by recombinations to the ground level 
can be captured by neutral atoms 
and produce additional Ly$\alpha$ radiation. The asymptotic yield in the 
extremely thick case
(case B approximation, where no continuum photon can leave the cloud) is 
$\epsilon_{\rm thick}=\alpha^{\rm eff}_{^2P}/\alpha_{\rm B}\sim 0.65$.
We use this value to compute the emission rate of fluorescent Ly$\alpha$ photons in the
simulation box.

\begin{figure}
\plotone{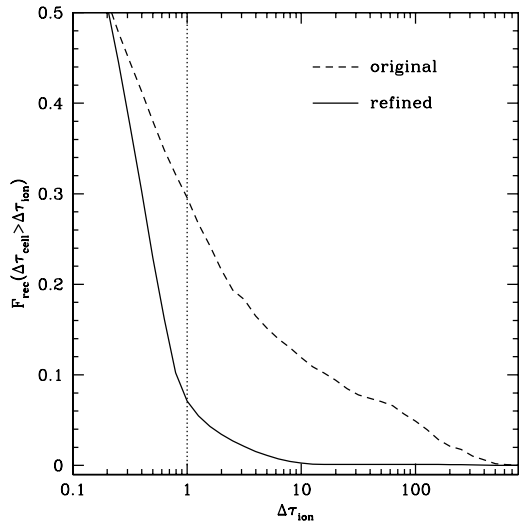}
\caption{The fraction of all hydrogen recombinations happening in the simulation box 
which take place in cells 
whith a single-cell HI optical depth (at the Lyman Limit) $\Delta\tau_{\rm cell}> 
\Delta\tau_{\rm ion}$ is plotted as a function of $\Delta\tau_{\rm ion}$.    
The function $F_{\rm rec}$ equals unity when  $\Delta\tau_{\rm ion}$ is equal to (or smaller than)
the minimum optical depth of a single cell in the simulation box.
Dashed and solid lines respectively refer 
to the original and the adaptively refined simulation boxes.
Note that a proper treatment of the radiative
transfer problem requires that the Ly$\alpha$ photons are generated
within cell with $\Delta\tau_{\rm cell}\lesssim1$ (see text).}
\label{rf}
\end{figure}

\subsection{Resolving the optical depth}

When we apply the method described above to our simulation, we find that 
the shielding layers 
(where the transition between optically thin and optically thick 
regions occurs) 
are poorly resolved (see Fig. \ref{rf}). 
Typically, they consist of very few cells
which each correspond to an HI optical depth variation (at the Lyman limit) of
 $\Delta\tau_{\rm cell}\equiv \Delta\tau(\nu_0)\gtrsim 1$. 
However,
for a proper treatment of the radiative transfer problem, more stringent
requirements on the grid spacing must be met. 
In particular, the Ly$\alpha$-emitting regions must be 
resolved with $\Delta\tau_{\rm cell}\lesssim 1$.
If not, both the spatial distribution of
recombinations and the escape probabilities of Ly$\alpha$ photons along 
different directions (see \S \ref{MC}) are spuriously altered.

To solve this problem, we adaptively refine the Ly$\alpha$ emitting regions by 
interpolating the original density and velocity 
fields of the input simulation. 
We use the solution of the radiative-transfer problem for the original
(unrefined) grid to select the regions to interpolate and the factor of 
refinement. 
Given the memory limitations of the available machines, 
we use a $100^3$ cells sub-box (which is particularly rich of structures)
of the original simulation
and we interpolate every cell with a significant recombination 
rate ($>0.1\%$ of the maximum) and $\Delta\tau_{\rm cell}>1$. 
The level of refinement is scaled proportionally to $\Delta\tau_{\rm cell}$ 
(up to a factor of 32 in each dimension) 
in order to have a subgrid of cells with 
$\Delta\tau_{\rm cell}\lesssim 1$.
Eventually, we re-compute the radiative transfer for the adaptively 
refined grid. Figure \ref{rf} 
shows that the fraction of recombinations originated in cells with 
$\Delta\tau_{\rm cell}>1$ decreases from 30\% to 7\% as a result of this 
refinement.
Moreover, in the finer grid, only a negligibly small number of recombinations 
takes place in 
extremely thick cells ($\Delta\tau_{\rm cell}>10$) compared with 12\% of the 
original grid.

As discussed in \S \ref{fcf}, we account for unresolved substructure 
in our simulation box by using
a non-vanishing clumping factor in the equation of ionization equilibrium.
Density variations within a parent cell of the original simulation due to
the refinement procedure described above could, in principal, significantly
contribute to the clumping factor. 
If this is the case, we should then adopt a value $\mathcal{C}<6$ 
for the refined
simulation to reproduce the observed abundance of LLSs.
We find that the clumping associated with the refinement is severe
in the densest zones of the simulation (which typically lie in the
self-shielded regions and do not contribute to the Ly$\alpha$ flux) 
but amounts to only a few per cent in the most rapidly recombining cells.
For these, we can then safely adopt $\mathcal{C}=6$ also for the refined box.

Increasing the spatial resolution of the simulation 
complicates the radiative transfer of ionizing radiation generated by recombinations.  
Equation (\ref{csi_ion}) assumes that every ionizing photon
generated by a HII recombination is absorbed in the same cell in which is generated.
However, this is no longer a good approximation for the adaptively refined cells which
are optically thin to UV radiation. 
In this case, ionizing photons generated by recombinations can be
absorbed in a different cell with respect to where they are created.
This process is too complicated to follow without an accurate radiative transfer scheme and
we use equation (\ref{csi_ion}) also for the refined cells.
How does this affect our results for the distribution of HI? 
First, the propagation of recombination radiation can slightly extend (of a few cells)
the thickness of the shielding layer of a gas cloud with respect to our results. 
The effect is probably more
pronuciated in the outer shells where the gas density is lower.
This should only redistribute the birth point of a small fraction of line photons.
On the other hand, in the central part of the shielding layer (which contributes most recombinations)
we expect that the flows of incoming and outcoming recombination radiation should nearly balance
given that the hydrogen density shows little variations.
In summary, our approximated treatment of recombination radiation should only slightly
modify the spectral energy distribution of the emerging Ly-$\alpha$ line

\subsection{Ly$\alpha$ radiative transfer}\label{MC}

We now combine
the results of the previous sections (namely, a set of arrays containing the Ly$\alpha$ emission rate, the HI density and the gas velocity field as a function of spatial position) to compute the spectra 
and the projected image on the plane of sky of the fluorescent sources.
The radiative transfer of resonant Ly$\alpha$ photons is modeled using a three-dimensional Monte Carlo scheme analogous to that employed by Zheng \& Miralda-Escud\'e (2002b, see also Ahn, Lee \& Lee 2001).
The method follows a large number of photon trajectories as 
they are scattered within the HI density and velocity distribution 
of the hydro-simulation. 

\subsubsection{Emission of Ly$\alpha$ photons}
We assume that Ly$\alpha$ photons are isotropically emitted with frequency $\nu_0$
in the frame of the recombining atoms (the natural linewidth is negligibly small for our purposes).
In the cosmic frame (e.g. for an observer lying at the center of the simulation box
and which participates to the free expansion of the universe),
the frequencies of the resonant photons appear
Doppler shifted by the projected velocities of the atoms along the photon trajectories.
The velocity of a hydrogen atom with respect to the cosmic frame is given by the superposition
of the Hubble flow with the bulk motion of the gas (i.e. the peculiar velocity of the fluid in the
corresponding cell of the simulation) and a random thermal velocity:
\begin{equation}
{\mathbf v}=H(z){\mathbf r}+{\mathbf v}_{\rm gas}+{\mathbf v}_{\rm th}
\end{equation}
with ${\mathbf r}$ the atom position with respect to the center of the simulation box.
The component of ${\mathbf v}_{\rm th}$ along the direction of the emitted photon
is generated by extracting
a  Gaussian deviate out of a distribution with zero mean and
dispersion $\sigma_{\rm th}=(k_{\rm B}\,T/m_{\rm H})^{1/2}=
12.8\, (T/2\times10^4\,{\rm K})^{1/2}\,{\rm km}\,{\rm s}^{-1}$ 
(with $k_{\rm B}$ the Boltzmann constant and $m_{\rm H}$ the atomic mass).

\subsubsection{Absorption}

The photon frequency can be conveniently expressed in terms of the variable
\begin{equation}
x= \frac{\nu-\nu_0}{\Delta}
\end{equation}
which measures the frequency shift from the Ly$\alpha$ line center in units of the Doppler width, $\Delta=\sqrt{2}\,\nu_0\, \sigma_{\rm th}/c$, where $c$ denotes the speed of light. 
The mean scattering cross section of Ly$\alpha$ photons in the fluid frame is
\begin{equation}
\sigma_{{\rm Ly}\alpha}(x)=\sqrt{\pi}\,f_{{\rm Ly}\alpha}\frac{c\,r_{\rm e}}{\Delta}\,H(a,x)
\label{cs}
\end{equation}
where $f_{{\rm Ly}\alpha}$=0.416 is the Ly$\alpha$ oscillator strength,
$r_{\rm e}=2.82 \times 10^{-15}$ m is the classical electron radius  and
\begin{equation}
H(a,x)=\frac{a}{\pi}\int_{-\infty}^{+\infty}{\frac{e^{-y^2}}{(x-y)^2+a^2}\,{\rm d}y}
\label{Vo}
\end{equation}
is the Hjerting-Voigt function.
For the relatively low-densities we are interested in, atomic collisions are not important and
the damping coefficient $a$ can be expressed in terms of the spontaneous decay
rate $\Gamma$ as
$a=\Gamma/(4\pi\Delta)=3.3\times 10^{-4}\, (T/2\times10^4\,{\rm K})^{-1/2}$.

We use equation (\ref{cs}) to determine the distance covered by each photon before it is scattered by
an atom. 
We first extract a random deviate, $R$, from an exponential distribution function and then
we integrate the product $n_{\rm HI}\,\sigma_{{\rm Ly}\alpha}(x)$ along the
photon direction of motion until the resulting optical depth equals $R$.
If the photon still lies within the computational volume, we select the velocity
of the scatterer. 
Note that,
in order to be able to absorb line radiation, an atom must have a
velocity component along the trajectory of the incoming photon, $v_\parallel$, 
which closely matches the Doppler shift. 
From equation (\ref{Vo}), it follows that, in the fluid frame, 
$x_\parallel=v_\parallel/(\sqrt{2}\,\sigma_{\rm th})$ is characterized
by the following probability distribution
\begin{equation}
{\mathcal P}(x_\parallel)=\frac{a}{\pi\,H(a,x)}\,\frac{e^{-x_\parallel^2}}{(x-x_\parallel)^2+a^2}\;.
\end{equation}
We use the method presented by Zheng \& Miralda-Escud\'e (2002b) to generate deviates
which follow this statistic.
The perpendicular component of the thermal velocity
in the scattering plane, $x_\perp$,
is then extracted from a Gaussian distribution with a temperature-dependent
dispersion as described above.
\begin{figure}
\plotone{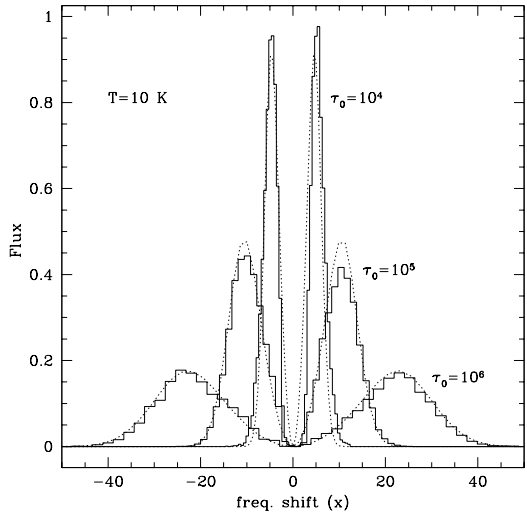}
\caption{Ly$\alpha$ spectrum emitted by a uniform slab with a midplane
source with optical depth $\tau_{0}$.
The results of our Monte Carlo code (solid histograms)
are compared with an analytical approximation (Neufeld 1990)
which becomes exact in the limit $\tau_0\to \infty$
(dotted lines).
A temperature of $T=10\,K$ is assumed.}
\label{MCtest}
\end{figure}

\subsubsection{Re-emission}

A new direction for the photon is then randomly selected according to a phase function, $P(\cos \theta)$ (with
$\theta$ the scattering angle), determined by atomic physics.
Resonant scattering has an isotropic angular distribution, $P=1$,
while wing scattering is characterized by the Rayleigh phase function, $P=3(1+\cos^2\theta)/4$
(Stenflo 1980).
We find that the two angular distributions give consistent outputs. 
All the results
presented in this work are obtained assuming isotropic re-emission.

To determine the new photon frequency,
we assume that the scattering process is coherent in the reference frame of the scatterer
(partially coherent scattering). 
This is appropriate when the excited atom undergoes no collisions before re-emission and
the radiative damping coefficient is small (Avery \& House 1968). 
Both conditions apply to Ly$\alpha$ radiation emitted by gas in
the typical conditions of the shielding regions in the intergalactic medium.
Once the scattering angle and the
photon velocity of the scatterer are specified, it is straightforward to 
compute the
frequency shift of the re-emitted photon in the fluid frame:
\begin{equation}
x=(x_{\rm in}-x_\parallel)+x_\parallel\cos\psi+x_{\perp}\sin\psi
\end{equation} 
where $x_{\rm in}$ is the frequency shift of the incoming photon and
$\psi$ is the angle between the 
direction of the incident photon and the direction of the scattering atom. 
A Lorentz transformation is finally used to compute the frequency shift in the 
cosmic frame. 

The set of calculations described above is iterated until the photon escapes the computational box.
\subsubsection{Ly$\alpha$ spectra}

To produce spectra (and broad-band images) of the fluorescent emitters,
we compute the surface-brightness of the computational box along the observer's line
of sight (hereafter, the $x$-axis). 
At each scattering, the probability that a photon will be re-emitted along this direction is
\begin{equation}
\frac{1}{4\pi}\,P(\cos\theta_x)\,e^{-\tau_x}
\end{equation}
where $\theta_x$ is the angle between the incoming photon and the $x$-axis and
$\tau_x$ denotes the Ly$\alpha$ optical depth of the scattering site
along the observer's line of sight.
\footnote{This optical depth includes the effects of
neutral hydrogen lying in the foreground of the computational box.}
For each photon and for each scattering,
we sum this quantity to a counter in correspondence of the projected position of the
scattering site and of the photon frequency. 
We thus obtain a three-dimensional array containing the
surface brightness of fluorescent Ly$\alpha$ photons as a function of 2 spatial coordinates plus frequency.
Note that
a simulated photon tends to remain for many scatterings in a rather small region before it eventually
escapes. This means that photons contribute only to a few pixels surrounding their
emission site.

Following Zheng \& Miralda-Escud\'e (2002b), we test 
our implementation of the Monte Carlo scheme against the analytical 
approximation by Neufeld (1990)  
for the optically thick, plane-parallel case.
Figure \ref{MCtest} shows that our code accurately reproduces 
the analytical solution which becomes exact in the limit of
extremely large optical depths.

\begin{figure*}
\epsscale{1}
\plottwo{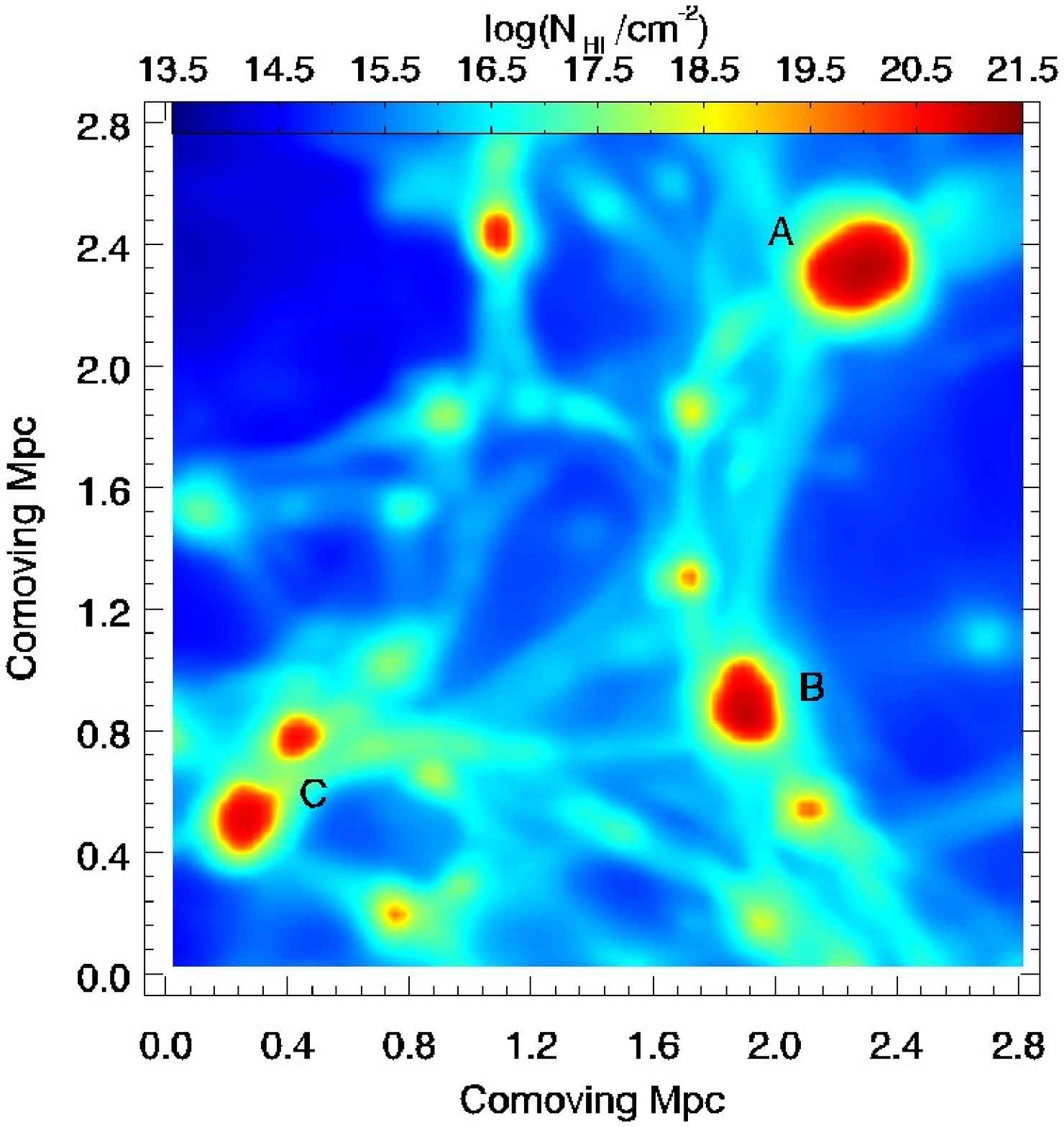}{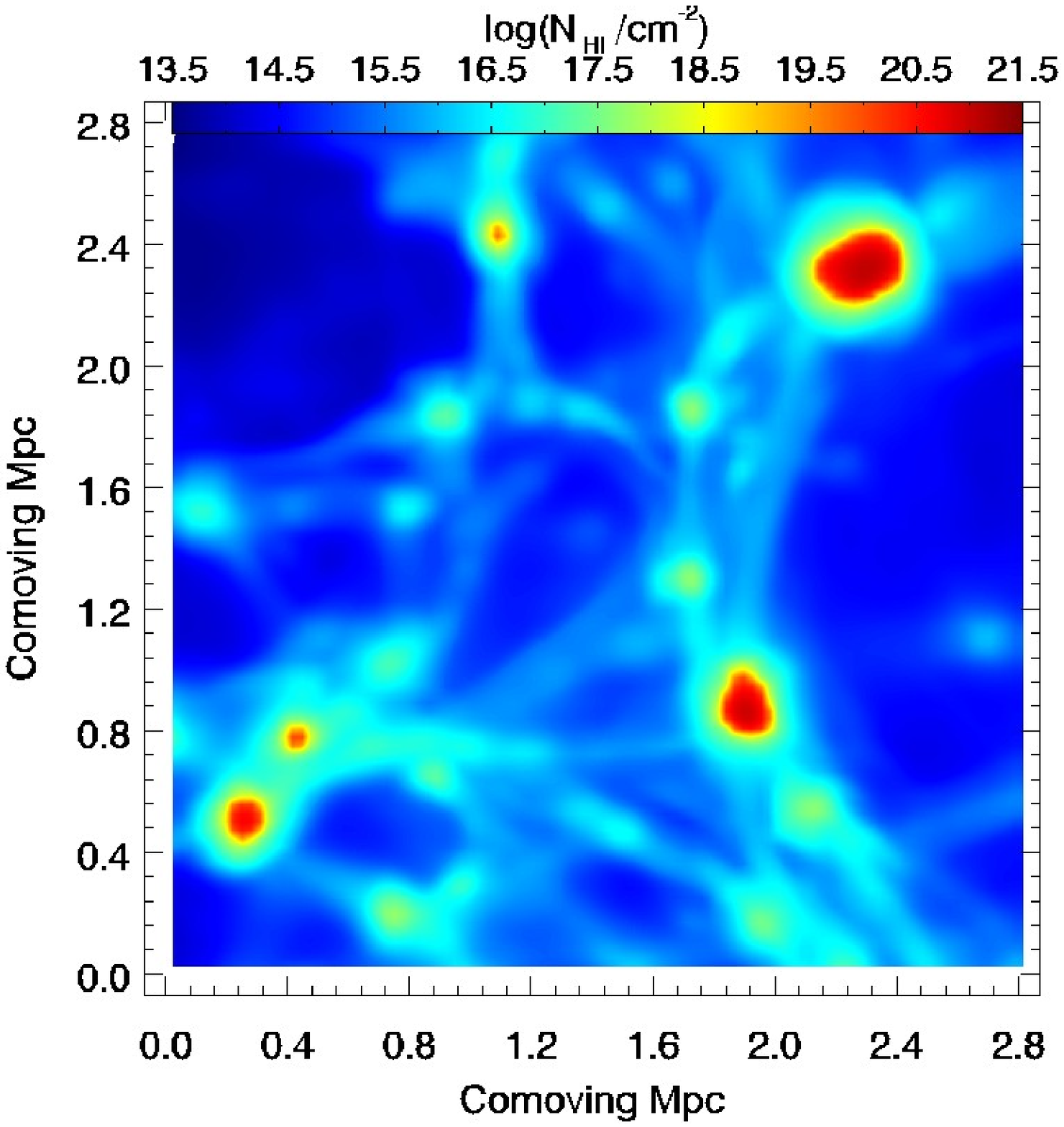}
\caption{Column-density distribution of neutral hydrogen at $z=3$.
In the left panel, the gas is
exposed to a diffuse UV background generated by the population of 
galaxies and quasars.
In the right panel, the ionizing flux from a foreground quasar,
located a short distance in front of the region and corresponding to
a boost factor $b=6$ (see equation (\ref{bdef}))
is superimposed to the diffuse background.}
\label{cden}
\end{figure*}
\begin{figure*}
\plottwo{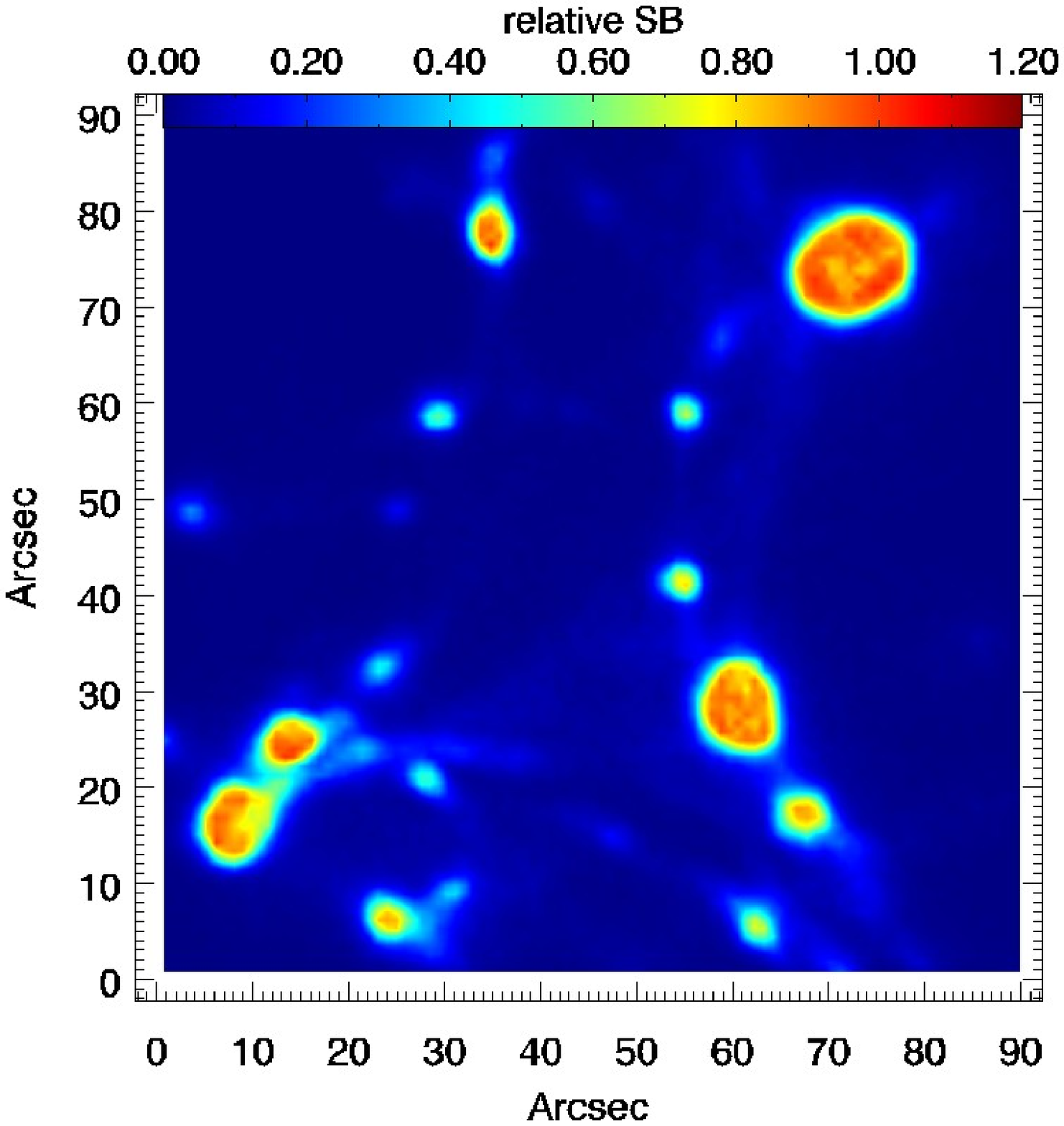}{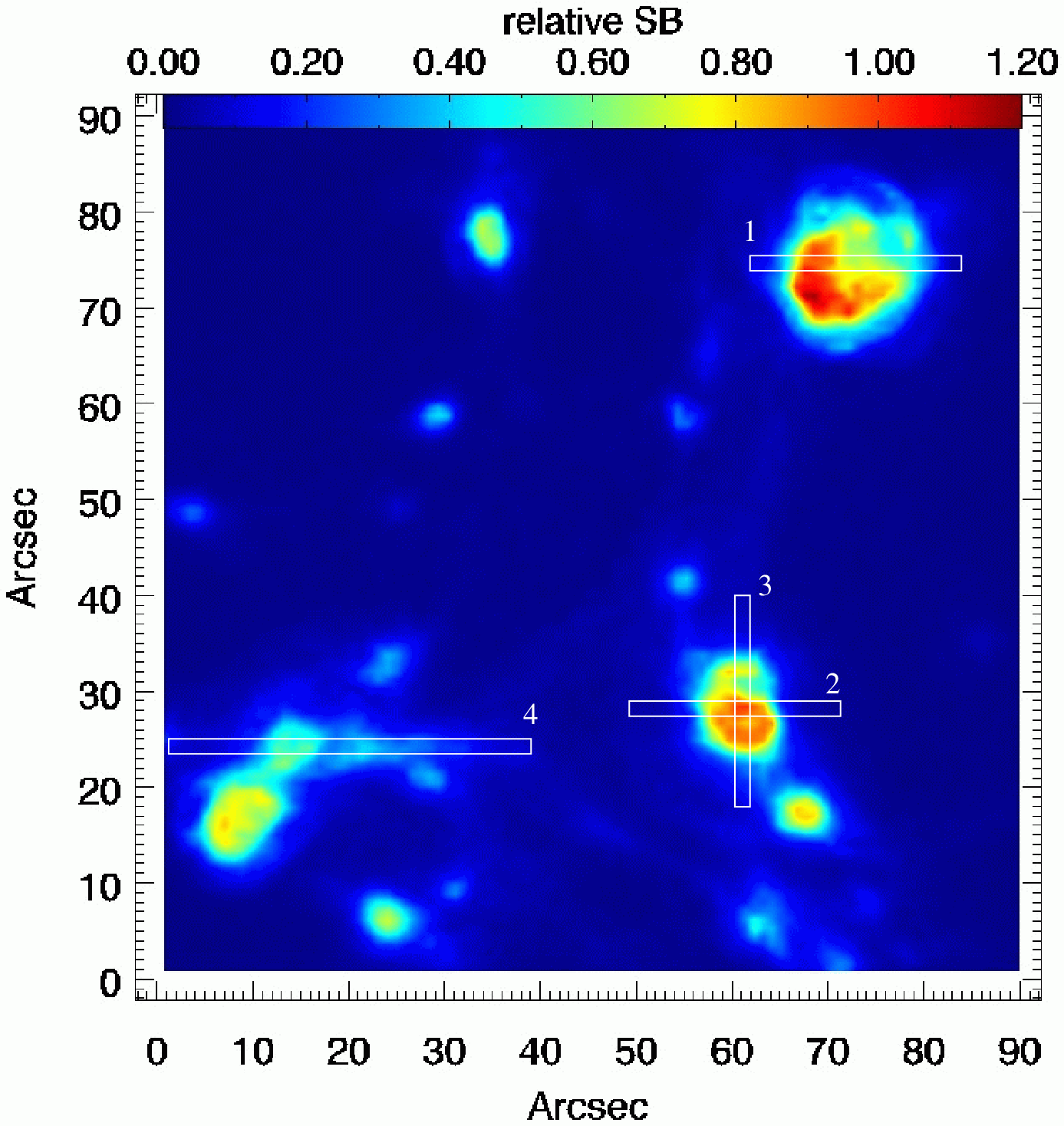}
\caption{Broad band images 
($\sim90\,$\AA, centered on $4864\,$\AA) of fluorescent Ly$\alpha$ emission
at $z=3$ for static gas clouds (left) and accounting for the gas velocity
field (right).
Both images correspond to the colum density distribution presented  
in the left panel of Figure \ref{cden}.
The white boxes indicate the location of the slit spectrographs
used to obtain the energy distributions presented in Figure \ref{spectra}.
\label{bbandHM}}
\end{figure*}

\section{Results}

In order to have an acceptable compromise between spectral resolution and CPU 
time, we only apply the Monte Carlo radiative transfer to the 
adaptively refined grid 
corresponding to a $100^3$ region of the original simulation box. 
To achieve a good signal-to-noise ratio, 
we generate $10^6$ photon trajectories for every simulation.
We thus obtain high resolution spectra for each pixel of
the resulting image that can be combined to simulate 
slit, line-emission integral field or broad-band observations.

\subsection{Diffuse background and static gas}
\label{dbsta}
We first discuss the ideal case of a static gas distribution 
illuminated with a uniform and isotropic backround of ionizing radiation.
This is obtained by artificially setting
to zero the velocity field of the gas within our refined box.

In the left panels of Figures \ref{cden} and \ref{bbandHM}, 
we respectively show the HI column density distribution and the
broad-band images ($\sim$ 90 \AA\ in the oberved frame, centered at 
$\lambda=(1+z)\cdot1216\, {\rm \AA}=4864$ \AA)  
of the selected region illuminated with the diffuse UV background. 
The color code in Figure \ref{bbandHM} gives the fluorescent 
Ly$\alpha$ emission 
rate (photons per unit time, surface and solid angle) 
in units of the impinging rate of ionizing photons times $\epsilon$ 
(i.e. the fraction of the recombinations yielding a Ly$\alpha$ photon):
\begin{equation}
R_{\rm HM}=\epsilon_{\rm thick}\int_{\nu_0}^{4\nu_0}
{\frac{J_{\nu}^{\rm HM}}{h_{\rm P}\nu}\mathrm{d}\nu}=2.44\times 10^{4}\
 \mathrm{cm}^{-2}\, \mathrm{s}^{-1} \, \mathrm{sr}^{-1}
\end{equation}
with $\epsilon_{\rm thick}\simeq0.65$.
\footnote{There is some observational evidence that the UV background at $z=3$ is dominated
by quasar emission with a negligible contribution from star-forming galaxies (e.g. Scott et al. 2000). 
In this case, the models by Haardt \& Madau (in preparation) give
$R_{\rm HM}=1.88 \times 10^4\ \mathrm{cm}^{-2}\, \mathrm{s}^{-1} \, \mathrm{sr}^{-1}$.
The spectral shape of the UV background  between $\nu_0$ and $4\nu_0$ 
is nearly identical to the general case discussed
in the main text. Therefore, our predictions for the surface brightness of fluorescent sources
can be simply scaled down by 30\% if future observations will prove that
galaxies do not significantly contribute to the ionizing
background at $z=3$. 
}
For an observer at redshift $z=0$, this corresponds to a Ly$\alpha$ surface 
brightness of
\begin{equation}\label{tsb}
{\mathrm{SB}}_{\rm HM}=3.67\times10^{-20}\  
\mathrm{erg}\,  \mathrm{cm}^{-2} \,\mathrm{s}^{-1} \, \mathrm{arcsec}^{-2}\;.
\end{equation}
The brightest fluorescent sources correspond to
compact gas clouds with a meatball topology. 
This is because the diffuse UV background is bright enough to fully ionize 
gas concentrations with $\rho \simlt 100\, \bar{\rho}$. In general,
the shielding regions either lie within virialized structures or correspond 
to dense gas shells which are accreting onto collapsed objects. 
As we will see below, the velocity field of the infalling gas 
produces specific signatures in the Ly$\alpha$ spectra.

The compact fluorescent sources lie along the filaments and sheets which 
characterize the
distribution of neutral hydrogen on cosmological scales.
For ease of reading, we label the three largest structures 
(which each have a diameter of $\sim0.4$ comoving Mpc) with the letters 
A, B and C (see Fig. \ref{cden}). 
Cloud C is composed of  two sub-units and is a part of an elongated structure 
which extends towards
cloud B. Similarly, a filamentary plume of gas bridges clouds A and B. 

\begin{figure}
\epsscale{1}
\plotone{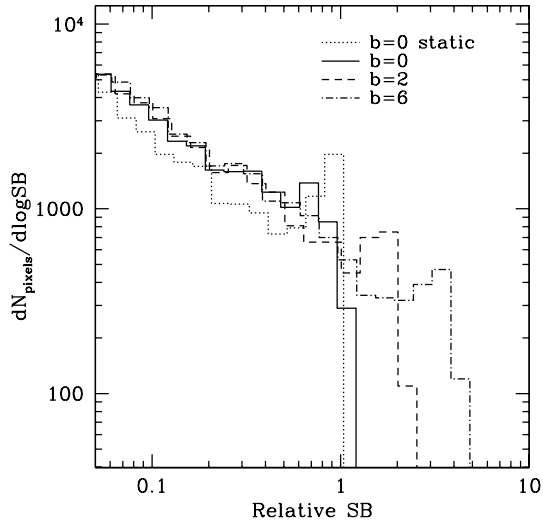}
\caption{Surface brightness distribution of fluorescent sources
ionized by a diffuse UV background (solid), by the additional contribution
of a quasar with ``boost" factor $b=2$ (dashed) and $b=6$ (dot dashed).
The dotted line is analogous to the solid one but is obtained by artificially
setting to zero the gas velocity field.}
\label{hist}
\end{figure}

Simple reasoning based on the plane-parallel model for line
transfer suggests that, 
in the absence of photon sinks (e.g. dust), 
self-shielded (isotropically-illuminated) objects 
should shine with a surface brightness 
of $\mathrm{SB}_{\rm HM}$ (Hogan \& Weymann 1987; Gould \& Weinberg 1996).
In our static simulation (Fig. \ref{bbandHM}, left panel), 
the SB of self-shielded objects closely matches
the predictions of this simple plane-parallel model.
The SB distribution in the simulation (dotted histogram in Fig. \ref{hist}) 
shows a narrow peak at this expected value.
In general, the SB scales proportionally to $N_{\rm HI}^{1/2}$ 
in the optically thin regions and asymptotically approaches its maximum value 
for self-shielded objects 
(see the top-left panel in Fig. \ref{cdsb}).
The brightest lines of sight
in fluorescent Ly$\alpha$ 
correspond to optically thick systems with
column densities $N_{\rm HI}\gtrsim 10^{18}\ \mathrm{cm}^2$ 
which are thus associated with LLSs and DLAs. 
All the photons of the ionizing background 
are converted into Ly$\alpha$ radiation within
the shielding layers of these optically thick systems. 
In the absence of other sources of ionizing radiation, 
it is impossible to produce a stronger Ly$\alpha$ flux.
This explains why the brightest objects  
in the left panel of Figure \ref{bbandHM} have a uniform SB and 
sharp boundaries which correspond to the regions with  
$N_{\rm HI}\simeq 10^{18}\ \mathrm{cm}^2$ in the left panel of Figure \ref{cden}.

\subsection{Diffuse background and realistic gas velocities}
\label{dbvel}
We are now ready to discuss the more realistic case where we include the gas 
velocity field of the hydro-simulation.
The corresponding  Ly$\alpha$ emission rate is shown in the right panel
of Figure \ref{bbandHM}.
The overall pattern is similar to the static case, but a number of striking
differences are noticeable. Namely:
{\it i)} the SB of self-shielded objects is no longer uniform 
(e.g. the right-hand side of 
Cloud A is nearly a factor of 2 fainter than the left-hand side);
{\it ii)} the boundaries of the emitting regions are less sharp 
and self-shielded objects are surrounded by large, low-SB halos;
{\it iii)} self-shielded objects can be significantly fainter 
(or, very rarely, brighter) than in the static case. 
\begin{figure}[t!]
\epsscale{1}
\plotone{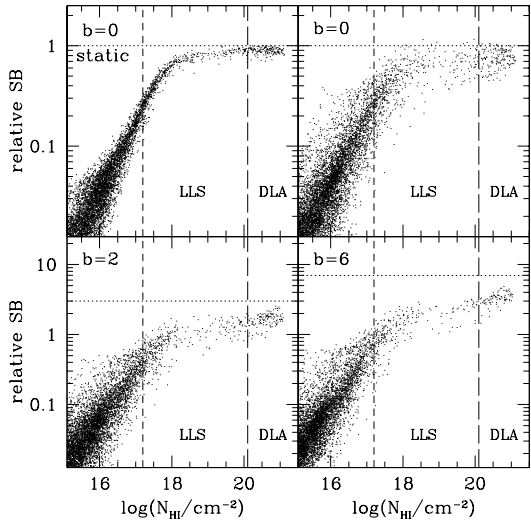}
\caption{The Ly$\alpha$ surface brightness of each pixel of the simulated
images is plotted against the corresponding column density of neutral
hydrogen. In the top panels, the intergalactic medium is ionized
by a diffuse background. In particular,
the top-left frame refers to a static gas distribution.
In the bottom panels, a quasar with boost factor $b=2$ (bottom-left) 
and $b=6$ (bottom-right) is superimposed to the diffuse background.
Dotted lines mark the expected SB for a plane-parallel slab while
dashed lines indicate the minimum column density for LLSs (short-dashed)
and DLAs (long-dashed).}
\label{cdsb}
\end{figure}

{\begin{figure*}[]
\plotone{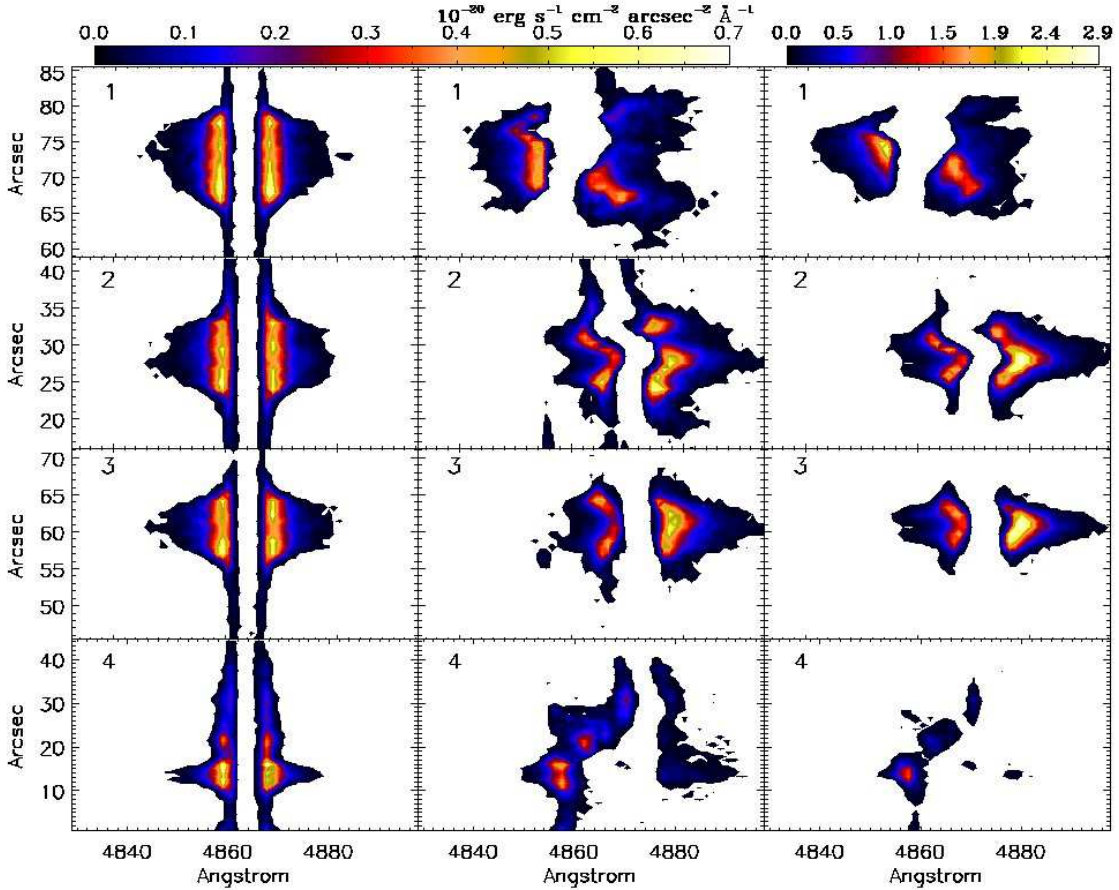}
\caption{Two-dimensional spectra obtained ``observing'' our simulations
with the slits shown in the right panel of Figure \ref{bbandHM}.
Different columns refer to different simulations. In particular, from left
to right: diffuse background and static gas distribution
(\S \ref{dbsta}), diffuse background and realistic gas velocities
(\S \ref{dbvel}),
quasar with $b=6$ plus diffuse background (\S \ref{dbqua}).
The labels in the top left corner indicate a particular slit spectrograph
as in Figure \ref{bbandHM}.
\label{spectra}}
\end{figure*}

The top-right panel 
in Figure \ref{cdsb} shows that the gas velocity field introduces additional
scatter into the SB - $N_{\rm HI}$ relation with respect to the static case. 
The brightest lines of sight
still correspond to $N_{\rm HI}\gtrsim 10^{18}\ \mathrm{cm}^2$
but now two regions with the same column density can be associated with
brightnesses which differ up to a factor of 5.
In consequence, the SB distribution 
of optically thick regions is broader and it is slightly shifted to fainter
fluxes with respect to the static case 
(see the peak of the solid histogram in Fig. \ref{hist}).
We find that the median SB of the self-shielded objects amounts to
nearly $75\%$ of the value predicted by Gould \& Weinberg (1996). 
At the same time, a larger fraction of the sky has
$\mathrm{SB}<\mathrm{SB}_{\rm HM}$ compared to the static case.
(the power-law part of the solid histogram in Fig. \ref{hist}).
As we will show below, this excess is caused by foreground scattering
of the Ly$\alpha$ photons and is related to the presence
of extended Ly$\alpha$ halos around self-shielded objects.

A better understanding of the ``velocity-field effect'' can be achieved
by comparing the spectra of the fluorescent emission in
the static and in the general case.
In the left and central panels of Figure \ref{spectra}, we show 
the corresponding spectral energy distributions
of the Ly$\alpha$ photons. These have been obtained 
positioning four slit spectrographs
(width $\simeq 0.9$ arcsec and variable length) 
on top of the three brightest sources 
as shown in the right panel of Figure \ref{bbandHM}.
In a static gas distribution, spectra have a characteristic double humped 
shape and are symmetric with respect to the line center. 
On the other hand, in the general case the energy distribution 
is no longer symmetric. 
In fact,
particular configurations of the velocity and density fields are able to 
strongly suppress one of the wings 
of the Ly$\alpha$ line and significantly lower the observed SB of the 
self-shielded objects.
In the particular case of Cloud A,
a low-density concentration of neutral 
hydrogen is infalling onto the Ly$\alpha$ emitting region.
The relative 
velocity (along the line of sight) corresponds to $\sim 4\sigma_{\rm th}$
and thus to a very high optical depth. 
Therefore, most of the photons that, in the static case,
leave the shielding layers along the line of sight in the red Doppler wing 
will scatter within the infalling cloud and escape in other directions
loweing the observed surface brightness.
These photons will then form the extended Ly$\alpha$ halos which surround
the brightest objects in Figure \ref{bbandHM}.
The phase-space distribution of
neutral gas in the vicinity of the emitting regions 
thus plays a fundamental role in reshaping the Ly$\alpha$ spectral 
energy distribution.
In broad terms, infalling material diminishes the red
wing of the spectrum, while gas which is receding from the emitting region
(which could also mean that
the shielding layer is infalling onto a central object
more rapidly than the surrounding gas) damps out the blue peak of the spectrum.
We find that
the velocity dispersion of the gas within the regions crossed
by the Ly$\alpha$ photons broadens the red and blue peaks of the spectrum 
by up to $10$ \AA. 
On the other hand, when both peaks are detectable, 
their separation is nearly independent from
the detailed properties of the emitting regions and is set by the
thermal velocity dispersion of the original cloud. 
In fact, in analogy with the plane-parallel case,
the escape probability of a Ly$\alpha$ photon peaks at a frequency which 
only depends 
on the optical depth of its emission site and on the temperature of the medium.
For a typical self-shielded cloud ($\tau_{\rm LL}\simgt 1$),
the spectrum peaks at  $\sim \pm4 x$  which, for the assumed temperature, 
corresponds to a separation of $\sim 8$ \AA.

The two-dimensional spectra shown in the central panel of Figure \ref{spectra}
clearly show that the gas velocity field within and in the vicinity of 
the shielding layers  
has a complicated structure which does not show the
characteristic pattern of ordered rotation or symmetric infall 
considered by Zheng \& Miralda-Escud\'e (2002b).

\begin{figure*}[]
\plottwo{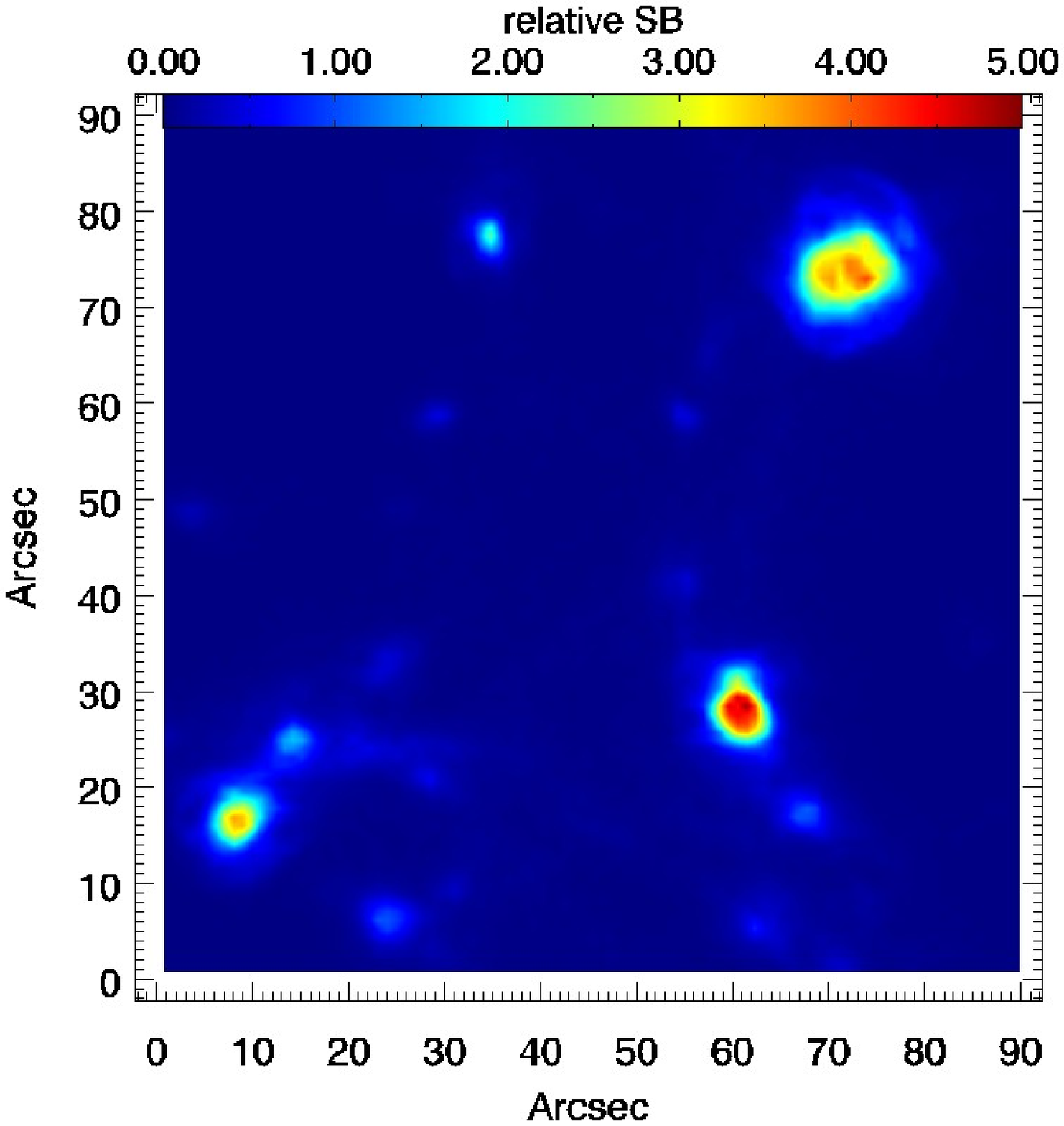}{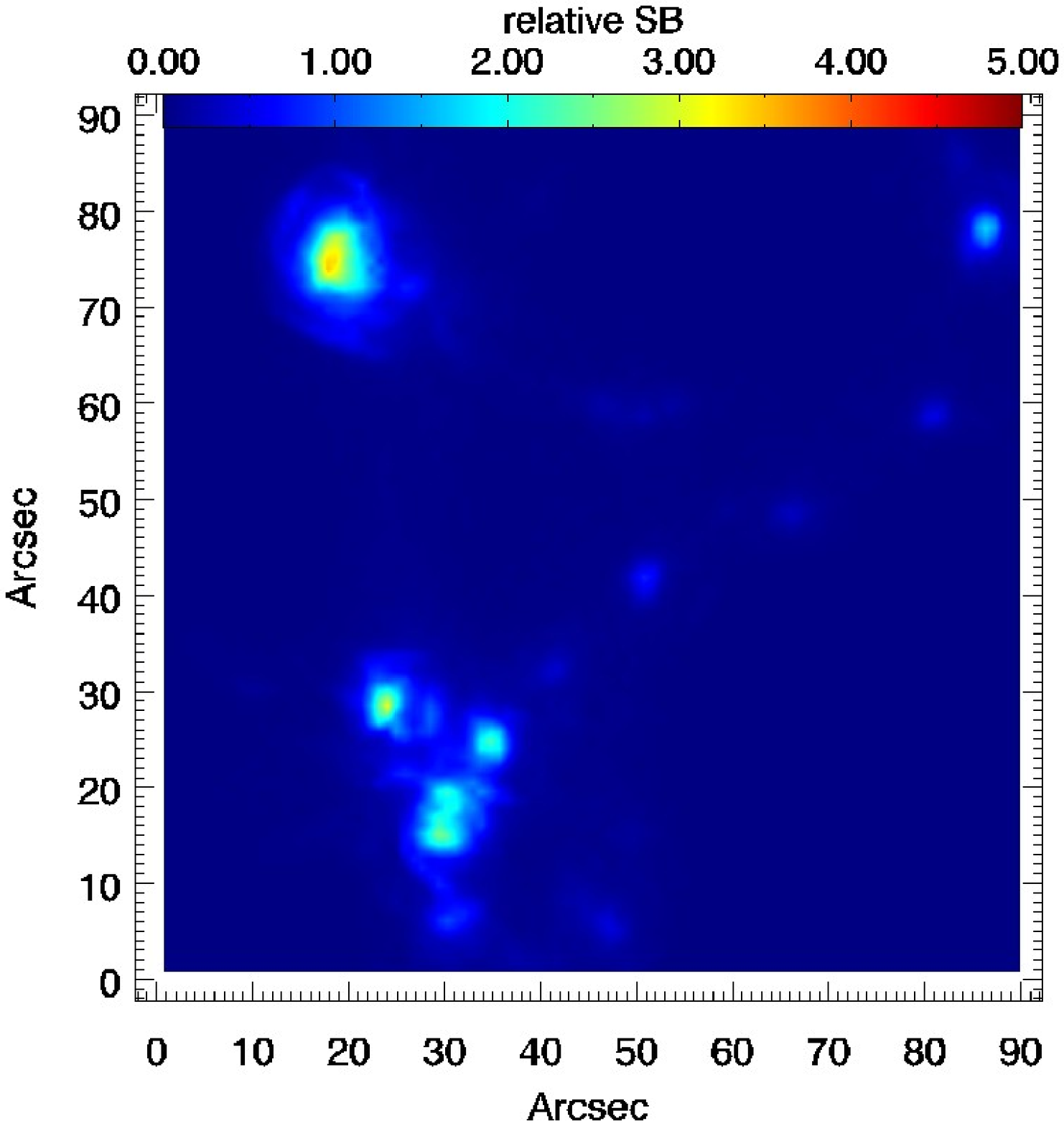}
\caption{Broad band images ($\sim90\,$\AA, centered on $4864\,$\AA) 
of fluorescent Ly$\alpha$ emission from our simulation box
at $z=3$.
In both cases, the intergalactic medium 
is ionized by a diffuse background and by a quasar with boost factor
$b=6$.
The image on the left (right) is obtained by observing our simulation box from
a line of sight parallel (perpendicular)
to the direction of quasar illumination.
The left frame corresponds
to the column density distribution presented in the 
right panel of Figure \ref{cden}. 
\label{bbandQSO}}  
\end{figure*}}

\subsection{Quasar plus diffuse background}
\label{dbqua}

We now discuss a case of anisotropic illumination, 
obtained by superimposing the ionizing flux from a quasar
to the diffuse UV background. 
The quasar is imagined to lie a short distance in front of 
the computational box as seen by us,
and thus enhances the UV illumination experienced by faces of gas clouds
exposed to it.
Note that  the ``boost'' factor $b$ (defined in \S \ref{RTran}) 
is determined by the intrinsic luminosity of the quasar 
and by its actual separation from the simulated region. 
The definition above can be generalized to any given
quasar spectrum using the emitted rate of ionizing photons.
At a physical distance $r$ from a quasar
with monochromatic luminosity $L_\nu=
L_{\rm LL}(\nu/\nu_{\rm LL})^{-\alpha}$, we find
\begin{equation}\label{bdef}
b=15.2 \,\frac{L_{\rm LL}}{10^{30}\, \mathrm{erg}\, \mathrm{s}^{-1} \, 
\mathrm{Hz}^{-1}} \,
\frac{0.7}{\alpha}\,\left(\frac{r}{\mathrm{1\,Mpc}}\right)^{-2}\;.
\end{equation}
The resulting $N_{\rm HI}$ distribution (assuming a boost factor $b=6$)
is shown in the right panel of Figure \ref{cden}.  
The corresponding broad-band image (obtained accounting for gas velocities)
is presented in the left panel of Figure \ref{bbandQSO}.
As expected, the self-shielded regions 
(and thus the fluorescent sources) are smaller
with respect to the isotropic background case
due to the extra-ionizing radiation produced by the quasar.
This also makes the fluorescent sources brighter (dot-dashed histogram in 
Fig. \ref{hist}) 
since more recombinations will be produced to balance a stronger ionization rate.
Based on the (plane-parallel) slab model, 
where Ly$\alpha$ photons are emitted 
following a cosine law (Gould \& Weinberg 1996), 
one would have naively expected 
an increase in the Ly$\alpha$ surface brightness towards the observer 
by a factor $1+b=7$ with respect to the diffuse background case. 
\footnote{This holds for normal incidence. In general, the surface
brightness of a slab which forms 
an angle $\theta$ with the incident quasar flux 
corresponds to a factor $1+b \cos\theta$.}
However, Figures \ref{hist} and \ref{cdsb} clearly indicate 
that the slab model overstimates the SB of the self-shielded objects. 
This is not due to shadowing effects.
In fact, the attenuation of the quasar flux by diffuse gas lying in front
of the fluorescent clouds is generally negligible.
Comparing with a static simulation, we also 
find that gas motions can only explain
a small part of this discrepancy.
In fact, in the presence of a quasar, foreground scattering is reduced
due to the lower neutral fraction present in low density gas and broad-band
images tend to be more uniform than in the case of isotropic illumination.
On the other hand, 
the slab approximation no longer applies 
when the size of the shielding layers is comparable with the radius of a cloud.
In this case, Ly$\alpha$ photons produced at a particular point
leave the cloud with a different angular
distribution with respect to the plane-parallel case.
For approximately spherical clouds and in the presence of uniform illumination,
this effect is suppressed for symmetry reasons. However,
when the ionizing flux is anisotropic, the Ly$\alpha$ SB does depend
quite strongly
on the geometry of the shielding layers.

\begin{figure}
\epsscale{1}
\plotone{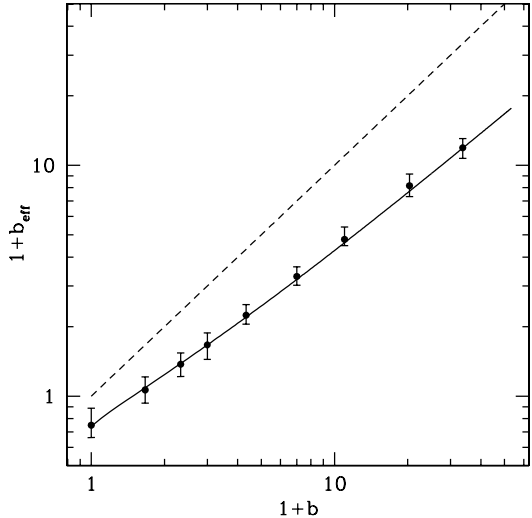}
\caption{
Ly$\alpha$ surface brightness of optically thick clouds
[expressed in terms of the effective boost factor defined in 
equation (\ref{beffdef})]
as a function of the impinging quasar ionizing flux 
[expressed in terms of the quasar 
boost factor, $b$, defined in equation (\ref{bdef})].
Points with errorbars denote the 25$^{\rm th}$, 50$^{\rm th}$ and 75$^{\rm th}$
percentiles of $b_{\rm eff}$ for the DLAs. 
The solid line represents the best-fitting relation given  
in equation (\ref{brel}).
Predictions of the static, plane-parallel model are plotted with a
dashed line.}
\label{bp}
\end{figure}

To study how the SB of self-shielded objects
along the quasar direction, $\mathrm{SB}_{\parallel}$, depends on
the impinging flux, we performed a series of simulations with increasing
$b$.
Our results are summarized in Figure \ref{bp}, where we
express  $\mathrm{SB}_{\parallel}$ in terms of an
``effective boost factor'' defined by
\begin{equation}
\mathrm{SB}_{\parallel}=(1+b_{\rm eff})\,{\mathrm{SB}_{\rm HM}}\;.
\label{beffdef}
\end{equation}
Points with errorbars mark the $25^{\rm th}, 50^{\rm th}$ and $75^{\rm th}$ 
percentiles of 1+$b_{\rm eff}$ among the DLAs. 
The solid line represents the best-fitting relation
\begin{equation}\label{brel}
1+b_{\rm eff}=0.74+0.50\,b^{0.89} \;,
\end{equation}
while the dashed line shows the predictions of the slab model.
Note that the geometric effect becomes more and more important as $b$
is increased.

Where do the ``missing'' Ly$\alpha$ photons go?
In the right panel of Figure \ref{bbandQSO}, we show
the fluorescent emission along a line of sight 
perpendicular to 
the direction of quasar illumination (assuming $b=6$ as in the left panel). 
In this case, the plane-parallel model predicts that the self-shielded
objects should emit at SB$_{\rm HM}$. In our simulations, however, 
the shielding layer deeply penetrates in the clouds along the quasar direction
and the slab model does not apply.
In consequence, self-shielded objects are much brighter than a slab along
this line of sight. 
Typically, $\mathrm{SB}_{\perp}\simeq
0.5\, \mathrm{SB}_{\parallel}$ for $b\gg 1$ while
$\mathrm{SB}_{\perp}\simeq \mathrm{SB}_{\parallel}$
for $b\ll 1$.
In other words, Ly$\alpha$ photons generated by the quasar
ionizing flux are emitted within a wide solid angle. As a consequence of this 
partial isotropization, self-shielded clouds are fainter than expected
(based on the slab approximation)  
along the quasar direction and brighter in the perpendicular directions.

Finally, 
in the bottom panels of Figure \ref{cdsb},
we show the SB - $N_{\rm HI}$ scatterplot for anisotropic illumination
(when observer, quasar and the simulation box are aligned).
It is worth noticing that, while the SB keeps nearly constant for LLSs,
on average, it steadily increases with $N_{\rm HI}$ for DLAs. 
This phenomenon can be explained by as follows.
Let us assume that self-shielded objects are nearly spherically symmetric.
Then,
i) the ionizing flux from the quasar depends on the incident
angle with respect to the local density gradient in the clouds;
ii) this cosine approaches unity for the central projected regions
of self-shielded objects;
iii) the column density reaches the highest values 
along these lines of sight.

\subsection{Size distribution of Ly$\alpha$ sources}

\begin{figure}
\plotone{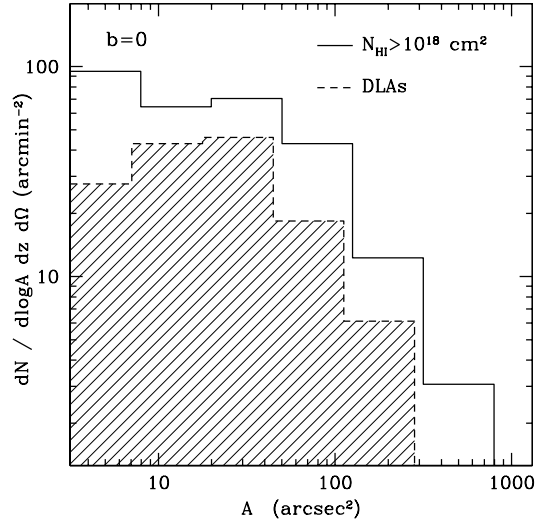}
\caption{
Differential size distribution of optically thick clouds (solid histogram) 
and DLAs (shaded histogram) in a simulation where the intergalactic
medium is exposed to a diffuse ionizing background.
The shaded histogram has been slightly shifted in the horizontal
direction to improve readibility.}\label{sd1}
\end{figure}

Knowing the size distribution of fluorescent Ly$\alpha$ sources is 
fundamental to planning an observational campaign for their detection.
Regrettably, our refined box is too small 
(its size being $\sim 2\, h^{-1}$ comoving Mpc) to provide a statistically 
representative sample of optically thick sources. On the other hand,
performing the line transfer on the $10\, h^{-1}$ Mpc box would require
an excessive amount of computer time. For these reasons, we decided to 
propagate only the ionizing radiation through the $10\, h^{-1}$ Mpc box
and to use the  
scatterplots in Figure \ref{cdsb} to convert the neutral-hydrogen column 
densities into Ly$\alpha$ fluxes.
In fact, independently of the value of $b$,
all lines of sight with $N_{\rm HI}>10^{18}\, \mathrm{cm}^{-2}$
are approximately associated with a constant Ly$\alpha$ surface brightness
(within a factor of 2 uncertainty caused by the gas motion and cosine
effects discussed above).
We then adopt this threshold value 
to derive the size distribution of fluorescent objects.
In Figure \ref{sd1}, we present our results for 
an isotropic ionizing background ($b=0$).
Solid and dashed histograms respectively refer to objects 
with $N_{\rm HI}>10^{18}\, \mathrm{cm}^{-2}$ and to DLAs.
It is worth remembering that we fixed the value of the clumping factor in our
simulation so as to reproduce the observed sky covering factor of LLSs. 
In consequence,
if a significant fraction of the real systems have a characteristic size which
is smaller than our numerical resolution, our simulation will overpredict
the number of large systems in order to preserve the required normalization.

\begin{figure}
\plotone{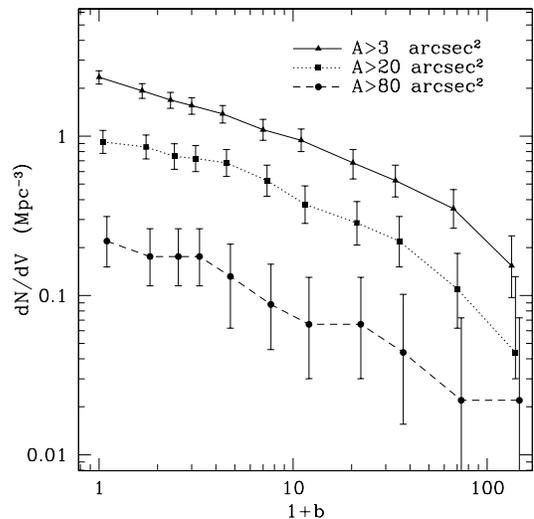}
\caption{Physical number density of fluorescent Ly$\alpha$ sources (with
size $\mathrm{A}$ indicated by the labels)
as a function of the impinging quasar flux [expressed in terms of
the boost factor $b$ defined in equation (\ref{bdef})]. Errorbars
are derived assuming Poisson statistics. Triangles and circles
have been slightly displaced in the horizontal direction to improve
readibility.}
\label{sd2}
\end{figure}

In Figure \ref{sd2}, 
we plot the number density of self-shielded objects as a function of $b$.
We use three different thresholds for the source size: 3 
(which corresponds to barely resolved objects), 
20 and 80 $\mathrm{arcsec}^2$.
In all cases, the number of sources rapidly drops with increasing $b$.
In fact,
higher values of $b$ characterize
regions which are closer to a given quasar (see eq. (\ref{bdef}))
and, obviously, correspond to a lower number density of self-shielded objects.

From this figure, it is also possible 
to determine the number density of sources which are brighter 
than a certain threshold value $(1+b_{\rm min})\,\mathrm{SB}_{\rm HM}$. 
Let us consider a Ly$\alpha$ source which is optically thick to
ionizing radiation at a given distance from a quasar.
Let us also imagine that we can move the cloud towards the quasar
thus increasing the $b$ factor.
As long as the cloud keeps optically thick, $\mathrm{SB}_{\parallel}$ 
monotonically increases. However, there exists a particular value
of the boost factor, $b_{\rm ss}$, 
at which the cloud is no longer able to self-shield.
Therefore, for $b\simgt b_{\rm ss}$, $\mathrm{SB}_{\parallel}$ keeps
roughly constant. 
\begin{footnote} {
The fraction of recombinations yielding 
a Ly$\alpha$ photon decreases from 
$\epsilon_{\rm thick}\sim0.65$ to $\epsilon_{\rm thin}\sim0.36$
when a cloud becomes optically thin. Therefore, we expect a fully ionized
cloud to be a factor of $\sim 2$ fainter in Ly$\alpha$
with respect to the optically thick case.}
\end{footnote}
Thus, the number of self-shielded objects at a given $b$ 
coincides with the number of sources (which are not 
necessarily optically thick)
with $\mathrm{SB}\gtrsim[1+b_{\rm eff}(b)]\,
\mathrm{SB}_{\rm HM}$.
In other words, the number of sources which are brighter than a given threshold
can be computed with the following procedure. First, convert the threshold
SB into an effective boost factor, 
$b_{\rm eff}^{\rm thr}$. Second, invert equation (\ref{brel})
to find the value of $b_{\rm min}$ such that 
$b_{\rm eff}(b_{\rm min})=b^{\rm thr}_{\rm eff}$. Third, use $b=b_{\rm min}$ in
Figure \ref{sd2} to determine the number density of the sources.
Fourth, use equation (\ref{bdef}) to find the volume within which 
it is possible to have $b>b_{\rm min}$.  

The variation of the number density of fluorescent sources as a function of $b$ is
somehow related to the proximity effect. Hydrogen clouds in the vicinity
of a quasar are strongly ionized and emit fluorescent radiation. In other words,
the missing absorption systems which determine the proximity effect can be detected
in emission through their recombination radiation. Therefore, in analogy with
studies of the proximity effect (e.g. Bajtlik, Duncan \& Ostriker 1988; Scott et al. 2000), 
the number density variation of fluorescent sources
around a quasar can be used to infer the intensity of the UV background.
In this case, reliable models of fluorescent emission are fundamental
to convert the observed counts into a background amplitude.

\subsection{Comparison with recent observational data}

We can use the above to 
compare the predictions of our models with the recent observational results
by Francis \& Bland-Hawthorn (2004, hereafter FBH).
These authors performed a deep narrow-band search for  
fluorescent Ly$\alpha$ emission in the vicinity of 
the $z=2.168$ quasar PKS 0424-131 ($L_{\rm LL}=1.67\times 10^{30}\,\mathrm{erg}\  \mathrm{s}^{-1} \mathrm{Hz}^{-1}$, 
$\alpha\simeq0.7$). 
At the 5$\sigma$ confidence level (corresponding to a surface brightness of
$4.7\times10^{-19}\,\mathrm{erg}\ \mathrm{cm^{-2}}
\ \mathrm{s^{-1}}\ \mathrm{arcsec^{-2}}$
for sources larger than $100\, \mathrm{arcsec}^2$
and to $9.6\times10^{-18}\,\mathrm{erg}\ \mathrm{cm^{-2}}$
for unresolved sources) no source was been detected. 
Based on the observed abundance of LLSs, FBH expected to find $\sim 6$
fluorescent clouds with a size of 100 arcsec$^2$. 
This estimate, however, 
does not take into account the ionizing radiation from the quasar.

Assuming that our results at $z=3$ are approximately valid at the quasar 
redshift, 
\begin{footnote}{We simply assume that the Ly$\alpha$ surface brightness 
scales as $(1+z)^{-4}$, i.e. 
$\mathrm{SB}(z=2.168)=2.54 \,\mathrm{SB}(z=3)$.}
\end{footnote} 
we find that the sensitivity limits of FBH correspond to
$b_{\rm min}\sim 11.2$ for sources which are larger than 100 arcsec$^2$.
Assuming that the ionizing backround keeps roughly constant in the redshift
interval $2<z<3$, from equation (\ref{bdef}) we find that this
corresponds to a distance from the quasar of 
$r_{\rm max}\sim 1.5$ (physical) Mpc. 
This is the maximum distance from the quasar at which an optically thick
cloud could have been detected.
Based on our simulations,
we expect to find, on average, 
$1-2$ objects within a sphere of radius 
$r_{\rm max}$ around the quasar. However, FBH limited their search
to distances smaller than 1 Mpc thus reducing the sampled volume by a factor
3.4 with respect to the theoretical limit. 
In this case, the expected number of sources ranges between 0.3 and 0.6.
Therefore, the probability of detecting (at least) one source
with one single observational run is $0.25<P<0.45$ (assuming Poisson
statistics).
Our simulations clearly show that the results by FBH 
are thus perfectly consistent with our understanding of the intergalactic medium
at high-$z$.
\begin{footnote}
{The detection limit for unresolved sources is less interesting.
It corresponds to $b_{\rm min}\sim 388$ (assuming that equation (\ref{brel})
can be extrapolated to such high values of $b$) 
and $r_{\rm max}\sim 250\,\mathrm{kpc}$. The number of expected sources
in the associated volume of $\sim 0.06\,\mathrm{Mpc}^3$ is therefore
negligibly small.}
\end{footnote}

There are caveats to the simple calculations descrived above.
For instance,
we have assumed that all the Ly$\alpha$ sources lying within a distance 
$r_{\rm min}$ from the quasar are brighter than $b_{\rm min}$.
This assumption tends to overestimate the number 
of detectable objects. In fact:
i) self-shielded clouds lying in front of the quasar are much fainter and
hardly detectable
(their SB actually depends on the angle between the line of sight and the
quasar direction);
ii) fully ionized clouds in the foreground of the quasar tend to be
a factor of $\epsilon_{\rm thick}/\epsilon_{\rm thin}\simeq 1.8$ fainter
than assumed above;
iii) if the age of the quasar is shorter than the hydrogen recombination
timescale no fluorescent sources will be detectable 
(see the discussion at the end of \S 2.2).
On the other hand, we have assumed that our simulation box is representative
of the gas distribution surrounding a quasar. 
Since optically selected quasars at high-$z$
tend to sit within the most massive dark-matter halos formed at that epoch 
(Porciani et al. 2004), it is reasonable to expect 
that matter clusters (and moves with larger peculiar velocities) around them. 
Therefore, we could have  
underestimated the number densities of fluorescent sources 
lying close to a quasar.

Despite of the approximations listed above, we believe that our results provide
the most reliable estimates for the abundance of fluorescent
Ly$\alpha$ sources at high-$z$ carried out to date.
Optimized sampling strategies are certainly required to observe these
objects. Our simulations then represent
a fundamental tool to plan observations around a given quasar.

\section{Uncertainties}
\label{discus}
\subsection{Radiative cooling}
While numerical simulations are a useful tool to guide our understanding, 
they cannot be considered a perfect model of reality.
A potential limit of our simulation is the lack of radiative cooling.
While this is not a concern for the diffuse intergalactic medium,
it becomes worrisome for highly overdense regions.
Fluorescent Ly$\alpha$ sources have intermediate overdensities 
($\rho/\bar{\rho}\sim 200$) and
are likely to be in equilibrium with the UV ambient radiation.
We thus expect them to be mildly affected by cooling processes.
In any case, most of the results discussed in this paper are nearly independent
of the details of the gas distribution.
Both the velocity and the geometric effects will be present anyway.
On the other hand, the size distribution of the sources might be more 
affected by the cooling processes.
It is also worth stressing that 
there is no way of accounting for the effects of cooling and heating
in a self consistent way. In fact, simulations where radiative
transfer is fully coupled with hydrodynamics are still not viable
with current supercomputers.
Moreover, other poorly understood processes (like
energy and momentum feedback) play an important role.
Therefore, it is hard to judge the level of approximation that current simulations
including gas cooling may reach.

\subsection{Resolution and sub-structure}
Is the finite resolution of our simulation affecting our results?
Multiple metal lines are often associated with single DLAs 
(e.g. Prochaska \& Wolfe 1997)
thus suggesting the presence of a clumpy medium.
Unresolved sub-structure in our simulation might reduce
the velocity effect and modify the outcoming spectra.
The adopted value of ${\cal C}$ implies that at least $1/6$ of the volume
is in dense clumps.
If these sub-structures have a diameter 
which is comparable to the cell size, we only expect a minor modification
of our results. In fact, the gas velocity in the simulation should closely 
approximate the motion of these clumps.  
The only effect is then a slight Doppler-shift of the entire Ly$\alpha$
spectrum of each cell.
The opposite case, where 
each cell contains a large number of small substructures (Abel \& Mo 1998), 
can be approximately discussed by considering an
additional contribution to the thermal velocity dispersion.
Assuming a value of 
$\sigma_{\rm th}\sim50\,\mathrm{km}\,\mathrm{s}^{-1}$
(corresponding to roughly half the virial velocity of the host halos,
Haehnelt, Steinmetz \& Rauch 1998),
we find that
the velocity effect is be strongly suppressed 
and the spectra are in better agreement with the slab model.
Note, however, that the existence of a sea of small subclumps
is disfavored by observations.
In fact, this scenario would produce broad absorption features instead
of the multiple metal systems associated with single DLAs
(cf. Haehnelt, Steinmetz \& Rauch 1998; McDonald \& Miralda-Escud\'e 1999).
In any case,
the velocity dispersion within our simulated DLAs 
is of the order of $100\,\mathrm{km}\,\mathrm{s}^{-1}$, in good agreement
with observational data. 

\subsection{Additional sources and dust}

Beyond fluorescent emission,
additional Ly$\alpha$ radiation might be produced in the inner regions of 
the clouds.
Within gravitationally collapsed objects, 
the gas tends to dissipate its internal energy by emitting 
line photons (e.g. Haiman, Spaans \& Quataert 2000; Fardal et al. 2001;
Furlanetto et al. 2005).
Similarly, internal star formation could act as a copious
source of Ly$\alpha$ photons, but, whereas fluorescent emission is expected to 
extend over several tens of kiloparsec, the Ly$\alpha$ emission from star 
forming region should be more concentrated near the centers of galaxies
(Furlanetto et al. 2005). 
We have focused here on the fluorescent emission generated by recombinations
and these extra sources of line photons are not considered in our analysis.
We will present a comprehensive model of Ly$\alpha$ emitters 
in a future paper.

At the same time we did not consider the destruction of Ly$\alpha$ photons
by dust grains.
Little is known about the dust properties within the intergalactic medium
at $z\sim 3$ even though there is some evidence for the presence of dust
in DLAs (Fall \& Pei 1993). Nevertheless, 
the associated absorption of fluorescent photons 
is likely to be minimal due to the 
relatively low $N_{\rm HI}$ of the shielding layer 
(e.g. Gould \& Weinberg 1996).
On the other hand, absorption is expected to be more severe for Ly$\alpha$
photons produced close to and within the star-forming regions
where dust is likely to be more abundant and the Ly$\alpha$ escape probability
is lower.
 
\subsection{The UV background}
\label{uvb}

Our results are based on the Haardt-Madau model for the UV background.
At $z=3$, the adopted model has an amplitude at the Lyman limit of
$J(\nu_0)=4.04 \times 10^{-22}$ erg $\mathrm{cm^{-2}}$ $\mathrm{s^{-1}}$ 
$\mathrm{sr^{-1}}\mathrm{Hz^{-1}}$ 
and corresponds to an hydrogen ionization rate $\Gamma=1.15 \times 10^{-12}\ 
\mathrm{s}^{-1}$.
Even though on the low side, this is consistent with recent
observational studies of the proximity effect, which give 
$J(\nu_0)= 7.0^{+3.4}_{-4.4}\times 10^{-22}$
erg $\mathrm{cm^{-2}}$ $\mathrm{s^{-1}}$ $\mathrm{sr^{-1}}\mathrm{Hz^{-1}}$
and $\Gamma=1.9^{+1.2}_{-1.0}\times 10^{-12}\ \mathrm{s}^{-1}$
(Scott et al. 2000).
Our results should then considered conservative, because they account for a UV background intensity
which is a factor of $1.75$ lower than current observational estimates. Note, however, that a different
normalization of the UV background
cannot change our predictions for the number density of fluorescent sources. 
In fact, we determined the clumping factor of the gas by imposing that the 
mean number of LLS in the simulation matches the observational data. Therefore, a stronger ionizing
radiation field would require a larger clumping factor to fit the observed counts.
If both $J_\nu^{\rm HM}$ and $\mathcal{C}$ are multiplied by the same multiplicative factor, $A_J$, 
the solution of the ionization equilibrium does not change. In fact, this is equivalent to multiply
both sides of equation (\ref{csi_ion}) by a constant factor.
Therefore our results for the number density and size distribution of fluorescent sources are robust with
respect to the amplitude of the UV background. Note, however, that the Ly$\alpha$ surface brightness
of optically thick cloud would be higher by a factor $A_J$ with respect to equation (\ref{tsb}).
Consistently, the boost factor in equation (\ref{bdef}) would decrease by a factor $A_J$ while
equation (\ref{brel}) would remain unaltered.

Some (most likely minor) 
modification of our number counts is instead expected if the spectral energy distribution
of the UV background significantly differ from the assumed one.
This crucially depends on the relative contribution of galaxies and quasars to the cosmic
ionizing background. 
Similarly, the spectral energy distribution of quasar radiation plays an important role.
For simplicity, we assumed that the quasar spectrum 
is identical to that of the cosmic background.
This corresponds to a energy index $\alpha\simeq 1.25$ i.e. to a much flatter 
spectrum than inferred from quasar observations ($1.5\simlt\alpha\simlt 1.8$).
What could be the effect of this assumption? 
The average mean free path of ionizing photons from a spectrum with $\alpha=1.8$ is only a factor of 
20\% smaller than for $\alpha=1.25$. 
Therefore, radiation from a steep quasar spectrum should penetrate a bit less
within hydrogen clouds with respect than our models. 
This might slightly modify the emerging Ly$\alpha$ spectra and
reduce the importance of the geometric effect. 
Note, however, that quasar radiation will be partially filtered by the IGM before reaching
the fluorescent clouds. Thus, a spectrum with $\alpha=1.8$ at emission will be transformed into
a flatter distribution in the shielding layer of a cloud.

\section{Summary}

We have presented a new method to produce realistic simulations of
fluorescent Ly$\alpha$ sources at high redshift.
We started by simulating the formation of baryonic large-scale structure in 
the $\Lambda$CDM cosmology.
A simple radiative transfer scheme was then used to propagate ionizing
radiation through the computational box and to derive the distribution of 
neutral hydrogen.
Finally, the transport of Ly$\alpha$ photons generated by hydrogen 
recombinations was followed using a three-dimensional Monte Carlo code.
As ionizing radiation,
we first considered the smooth background generated by 
galaxies and quasars. Then, as a second case, we superimposed to the background
the UV flux produced by a quasar lying in the vicinity of the simulation box.

Our detailed numerical treatment improves upon previous work which
was either based on rather crude approximations 
for the transfer of 
resonantly scattered radiation (Hogan \& Weynmann 1987; Gould
\& Weinberg 1996) or on highly symmetric semi-analytical models
for the gas distribution (Zheng \& Miralda-Escud\'e 2002b).
Our results show that simple models (e.g. Gould \& Weinberg 1996)
tend to overpredict the Ly$\alpha$ flux emitted from optically thick clouds. 
In fact, we identified two effects that  
reduce the fluorescent Ly$\alpha$ flux (and modify 
the spectral energy distribution)
with respect to the widespread static and plane-parallel model.

{\it Velocity effect --}
The velocity field inside and around the shielding layers 
of a gas cloud influences the emerging line profile.
The symmetry of the double humped spectrum is lost
and, in most cases, one of the two peaks is severely suppressed.
On average, the SB of a cloud is reduced by $25 \%$ with respect to 
the static situation.

{\it Geometric effect --}
For anisotropic illumination and in the presence of a strong
ionizing flux, the thickness of the shielding layer is
comparable to the size of the gas cloud.
In this case, the angular distribution of the emerging
radiation is very different than in the plane-parallel approximation.
For instance, close to a quasar, a cloud emits much less than 
predicted by the slab model 
in the direction of the quasar and much more in the other
directions.

The importance of these effects (in particular of the angular 
redistribution of Ly$\alpha$ photons) 
depends on the intensity of the impinging radiation.
In equation (\ref{brel}),
we provided a fitting function for the 
maximum Ly$\alpha$ brightness of optically thick sources
as a function of the incident ionizing flux.
In Figure \ref{sd2}, we presented our predictions for
the number density of fluorescent sources with different sizes. 

These results are consistent with the recent null detection by
Francis \& Bland-Hawthorn (2004) and represent a fundamental tool
for planning future observations.

\acknowledgements
We would like to thank Piero Madau for 
suggesting to consider the problem of anisotropic 
illumination in the presence of a quasar.
We benefited from extensive discussions with 
Francesco Haardt, Martin Haehnelt, Tom Theuns and 
all the participants to the ``Galaxy-Intergalactic Medium Interactions''
program recently held 
at the Kavli Institute for Theoretical Physics in Santa Barbara.
We thank the anonymous referee for useful suggestions which improved
the presentation of our work.
CP and FM acknowledge support from the Zwicky Prize Fellowship program at 
ETH-Z\"urich and SC from the Swiss National Science Foundation.

\begin{appendix}
\section{Radiative transfer for a diffuse background}
\label{6dtest}
The fluorescent Ly$\alpha$ flux from a source is proportional to the
number of hydrogen ionizations taking place in the 
corresponding gas cloud.
In order to simplify the radiative transfer problem for continuum radiation,
in equation (\ref{RT}) we only followed the propagation of ionizing
photons along the 6 principal directions of the simulation box.
In this appendix, we first provide justification for the multiplicative factor
$2\pi/3$ which appears in equation (\ref{RT}) and then test the
6-direction approximation.

\begin{figure*}
\plottwo{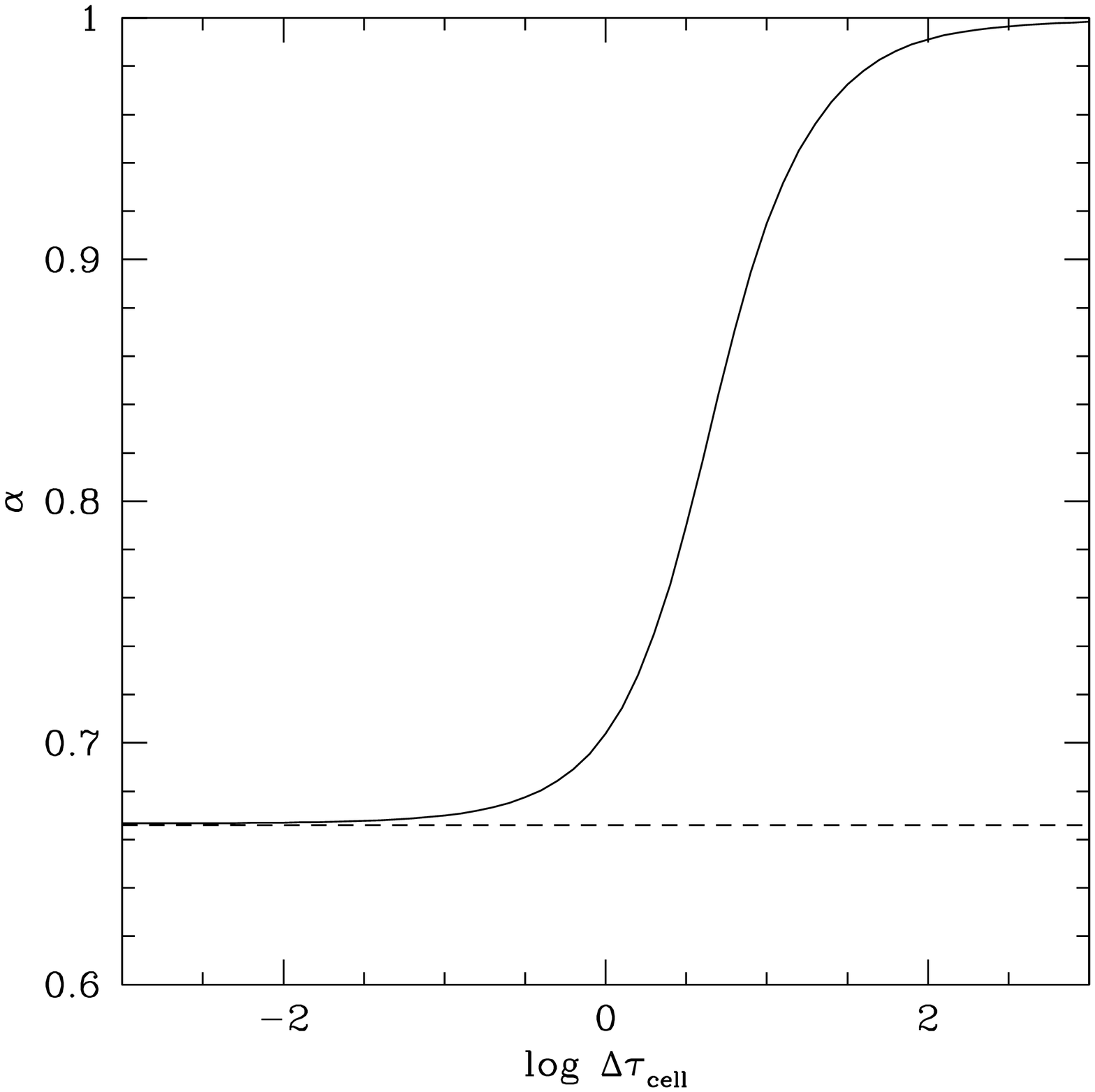} {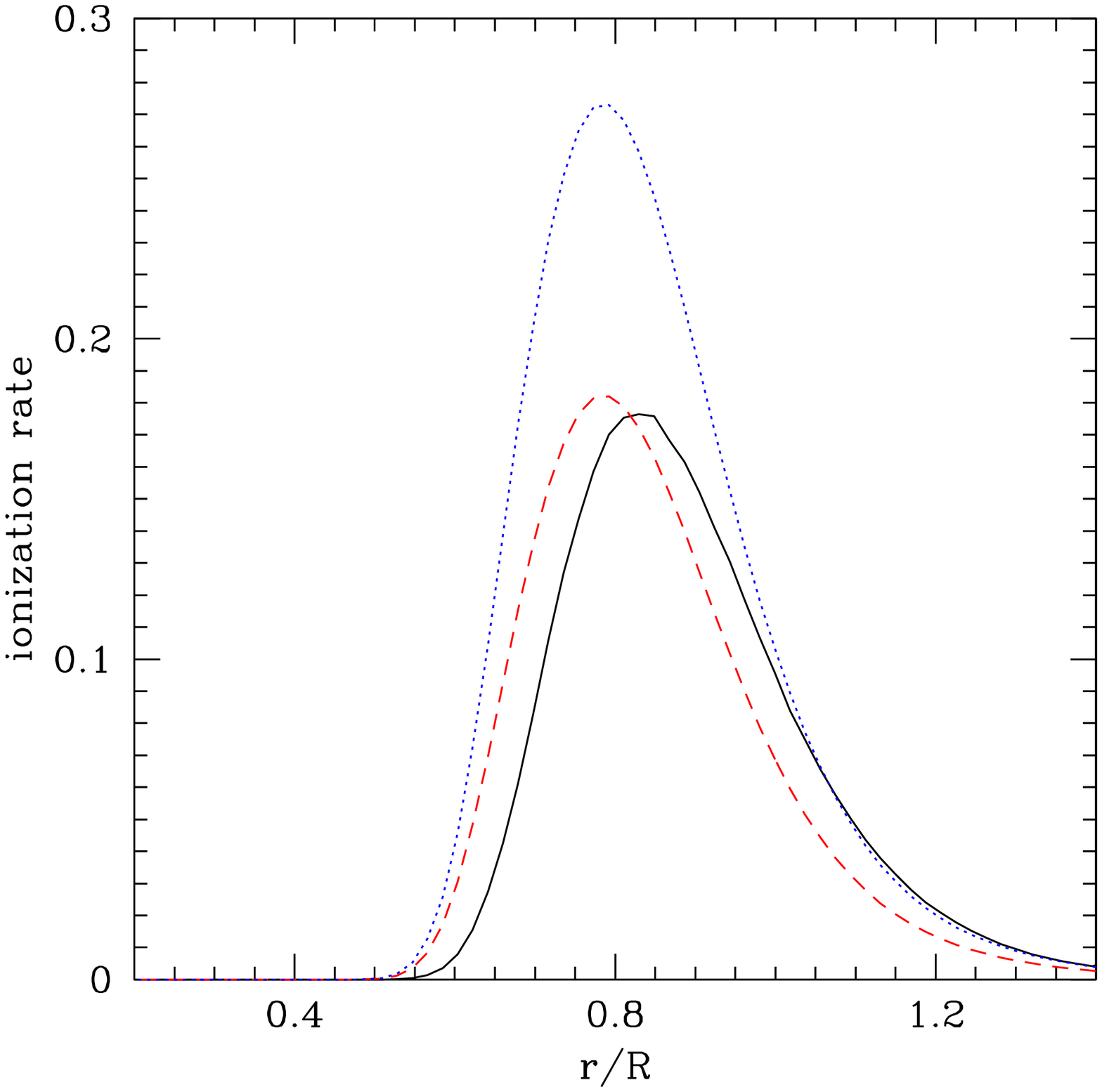}
\caption{
Left: The ionization rate generated by a uniform
UV background (of intensity $I$) penetrating in a cubic cell (of size $L$)
through one of its faces
as a function of the optical depth variation within the cell.
The solid line represents the ionization rate
(in units of  $\pi\,I\,L^2\,e^{-\Delta\tau_{\rm cell}}$)
while the dashed line marks the optically thin solution $\alpha=2/3$.
Right: Radial profile of the ionization rate (arbitrary units) for the
mock cloud described in Appendix \ref{6dtest}. The exact solution (solid)
is compared with our six-direction approximation in the optically thin (dashed)
and optically thick (dotted) cases.}
\label{cube2}
\end{figure*}

The factor $2\pi/3$ in equation (\ref{RT}) is obtained by discretizing
the left-hand side in equation (\ref{csi_ion}) along 6 preferred directions.
Note, however, than equation (\ref{csi_ion}) applies to an infinitesimal volume
element while our discretized version is used to describe finite cubic cells.
It is clear that the approximation is correct for optically thin cells, but
where does it break down?
Let us consider a homogeneous cube of side $L$ 
(corresponding to an optical depth $\Delta\tau_{\rm cell}$) 
illuminated by a uniform and monochromatic background with intensity $I$
(photons per unit time, surface and solid angle).
The ionization rate within the cube generated by the radiation background
impinging on one face
can be written as $\alpha\,\pi\,I\,L^2\,
e^{-\Delta\tau_{\rm cell}}$, with
$\alpha$ a numerical coefficient of order unity.
We use a Monte Carlo method 
to compute $\alpha$ as a function of 
$\Delta\tau_{\rm cell}$. Our results are plotted in Figure \ref{cube2}.
When the cell is optically thin $\alpha\simeq 2/3$, while 
$\alpha \to 1$ when $\Delta\tau_{\rm cell}\to \infty$. 
A rather sharp transition
between the two asymptotic regimes is observed for 
$\Delta\tau_{\rm cell}\sim 10$.
Note that, in the presence of a uniform background, the optically thin 
approximation adopted in the main text
is accurate even for $\Delta\tau_{\rm cell}=1$, where $\alpha=0.70$. 
Given that, in our adaptively refined cosmological simulation,
$\sim 93\%$ of recombinations take place in cells with
$\Delta\tau_{\rm cell}<1$, we used $\alpha=2/3$ in our calculations. 

All this, however, applies to an isolated cell exposed to the ionizing
background. 
We now want to test how accurately equation (\ref{RT}) works in the
shielding layer of an optically thick cloud. We thus consider
a spherically symmetric cloud which is optically thin at his external
boundary and thick at the center ($\log\Delta\tau_{\rm cell}=2-4\,(r/R)$, 
with $r$ the radial coordinate which vanishes at the cloud center and $R$
the characteristic size of the cloud, which roughly mimics one of the
clouds in our refined simulation).
We then illuminate the outer boundaries of the cloud with a uniform 
(over the external $2\pi$ sr) and monochromatic background.
This is injected from the surface of a cube of side $2R$ 
centered on the cloud and divided into $151^3$ mesh points.
In Figure \ref{cube2}, we compare the outcome of a detailed radiative transfer
code (Porciani \& Madau, in preparation) with our approximated method.
Note that in our six-directions approximation the ionizing flux penetrates
slightly deeper into the cloud with respect to the exact solution.
This happens because our method ignores photon trajectories which
are oblique to the principal axes of the box.
On the other hand, we find that our approximation produces 
92$\%$ of the ionizations that take place in the cloud.
Given the simplicity of our algorithm, this is a remarkable achievement.
Note that, adopting the optically thick approximation ($\alpha=1$), we would
have overestimated the number of recombinations by 38$\%$.
These figures are rather stable and, for realistic cases, 
do not depend on the details of the optical
depth distribution.

\end{appendix}


%

\end{document}